\documentclass[12pt,fleqn]{article}

\usepackage{amssymb}
\usepackage{epsfig}
\usepackage{eucal}
\usepackage{amsmath}

\numberwithin{equation}{section}

\newlength{\dinwidth}
\newlength{\dinmargin}
\renewcommand{\cal}{\CMcal}                  
\newcommand{\ps}{\!\cdot\!}                  
\newcommand{\ta}[1]{#1\hspace{-.42em}/\hspace{-.07em}} 
\newcommand{\pa}{\mbox{\scriptsize\sf a}}    
\newcommand{\pb}{\mbox{\scriptsize\sf b}}    
\newcommand{\pc}{\mbox{\scriptsize\sf c}}    
\newcommand{\pq}{\mbox{\scriptsize\sf q}}    
\newcommand{\pg}{\mbox{\scriptsize\sf g}}    
\newcommand{\as}{\alpha_s}                   
\newcommand{\ab}{\overline{\alpha}_s}
\newcommand{\kk}{{\boldsymbol k}}            
\newcommand{\ku}{{\boldsymbol k}_1}
\newcommand{\kd}{{\boldsymbol k}_2}
\newcommand{\qq}{{\boldsymbol q}}
\newcommand{\dif}{{\rm d}}                   
\newcommand{\du}{\dif[\ku]}
\newcommand{\dd}{\dif[\kd]}
\newcommand{\dk}{\dif[\kk]}
\newcommand{\ds}{\displaystyle}
\newcommand{\dq}{\delta[\qq]}                
\newcommand{\G}{{\cal G}}                    
\newcommand{\K}{{\cal K}}                    
\newcommand{\M}{{\cal M}}                    
\renewcommand{\P}{{\mathcal P}}              
\newcommand{\pp}{{\tt P}}                    
\renewcommand{\t}{{\boldsymbol t}}           
\newcommand{\f}{{\rm f}}                     
\newcommand{\e}{\varepsilon}                 
\newcommand{\po}{\epsilon}                   
\newcommand{\poc}{\overset{*}{\epsilon}}     
\renewcommand{\l}{\lambda}
\renewcommand{\o}{\omega}

\begin{document}

\title{\large\bf
 $\kk$-Factorization and Impact Factors at Next-to-leading Level
   \footnote{Work supported in part by the E.U. QCDNET contract
            FMRX-CT98-0194 and by MURST (Italy).}}
\author{M. Ciafaloni$^{\dagger}$ and D. Colferai$^{\dagger}$\\
{\em Dipartimento di Fisica, Universit\`a di Firenze} \\
{\em and INFN, Sezione di Firenze} \\
{\em Largo E. Fermi, 2 - 50125  Firenze}}
\date{}
\maketitle
\thispagestyle{empty}
\begin{abstract}
We further analyse, at next-to-leading $\log s$ level, the form of
$\kk$-factorization and the definition of impact factors previously
proposed by one of us, and we generalize them to the case of hard
colourless probes. We then calculate the finite one-loop corrections
to quark and gluon impact factors and we find them universal, and
given by the same $\K$ factor which occurs in the soft timelike
splitting functions.
\end{abstract}

\begin{center}
\end{center}
\vspace*{2.5 cm}
{\small $~^{\dagger}$ e-mail: ciafaloni@fi.infn.it, colferai@fi.infn.it}
\newpage


\setcounter{page}{1}
\section{Introduction}\label{intro}

The rise of small-$x$ structure functions at HERA \cite{a96} has
emphasized the need of understanding high-energy QCD \cite{bfkl77} at
next-to-leading $\log x$ (NL$x$) level \cite{fl89}-\cite{cc97}.

The general purpose of such calculations is to derive the NL$x$ corrections
to both the BFKL equation \cite{bfkl77} and to the resummed anomalous
dimensions \cite{cc97} in hard processes with one or more hard
scales. Therefore, we need to compute on one hand the NL$x$ BFKL kernel
\cite{fl98,cc98}, which is supposed to be independent
of the external probes, and on the other hand the impact factors
\cite{c98}, which characterize the probe.

The separation of impact factors and kernel is made on the basis of
high-energy $\kk$-dependent factorization (Sec.~(\ref{kfdp})), or in short
$\kk$-factorization \cite{cch90}, which is therefore to be extended at
NL$x$ level.

The purpose of the present paper is to further analyse a form of
$\kk$-factorization proposed by one of us \cite{c98}, by defining the
impact factors in a proper factorization scheme and by computing the
one-loop corrections to them in the case of partonic probes.

We have emphasized the use of a proper factorization scheme, because
the separation of the cross section in impact factor contributions and
gluon Green's function $\G_{\o}$ suffers of some ambiguity, analogous
to the one between coefficient functions and anomalous dimensions in
collinear factorization. In fact, the subtraction of the leading
$\log s$ terms involves a prescription which is not unique, not only
for choosing the scale of $s$, but also for the form of $\G_{\o}$ at
finite energies. Therefore, some probe-independent NL$x$ terms can be
attributed to either the impact factors or to the Green's function,
depending on the scheme being adopted.

A definition of impact factors on the  basis of a proper scale choice
for subtracting the leading term was given \cite{c98} for the case of dijet
production, and was later used \cite{cc98} in order to extract the
NL$x$ kernel \cite{fl98,cc98}. Here we further discuss this definition and the ensuing
factorization scheme in Sec.~(\ref{ifnla}). First, we emphasize the
requirements that the subtracted leading term should satisfy at finite
energies for the impact factors to be well defined (e.g., without
spurious infrared divergences).

Furthermore, we generalize the definition above to the case of
colourless probes, which is the most important one for applications to
DIS and to heavy quark processes. We argue on this basis that the NL$x$
kernel extracted before \cite{cc98} in a partonic process
applies to physical hard processes
as well, in the given factorization scheme.

We then proceed to the actual calculation of partonic impact factors
in Secs.~(\ref{qqcoll},\ref{ggcoll}). The latter still contain some
collinear divergences due to the massless initial partons, which are
those not subtracted out with the leading terms. We then factor them
out in proper splitting functions, and we end up with some finite
one-loop corrections not only for the quark \cite{c98}, but also for
the gluon initial state.

It is amusing to note that such finite renormalization factors
(Sec.~(\ref{cffp})) are universal, i.e. are the same for quark,
antiquark and gluon, and are given in terms of the coefficient
$\K=\left[N_c\left(\frac{67}{18}-\frac{\pi^2}{6}\right)-\frac{5n_f}{9}\right]$
in the $\kk$-factorization scheme of Ref. \cite{c98}. In other schemes
such finite renormalizations are different, but still independent of
the probe.

A similar property, and the same coefficient $\K$, were noticed to
occur in the one-loop corrections to the $1/(1-z)$ part of the
timelike DGLAP splitting functions. In this case the universality
feature is probably to be ascribed to soft gluon properties \cite{dmw96}
which appear to determine not only single parton emission, but
correlations as well. Apparently, a related argument can be applied to
gluon-Regge exchanges, along the lines suggested by Ref.~\cite{kk96}.

We finally discuss our results in Sec.~(\ref{discus}), by giving some
calculational details in Appendices A and B.


\section{k-Factorization in dijet production}\label{kfdp}

\begin{figure}[h]
 \centering
\begin{picture}(80,55)
  \put(0,0){\includegraphics[width=80mm]{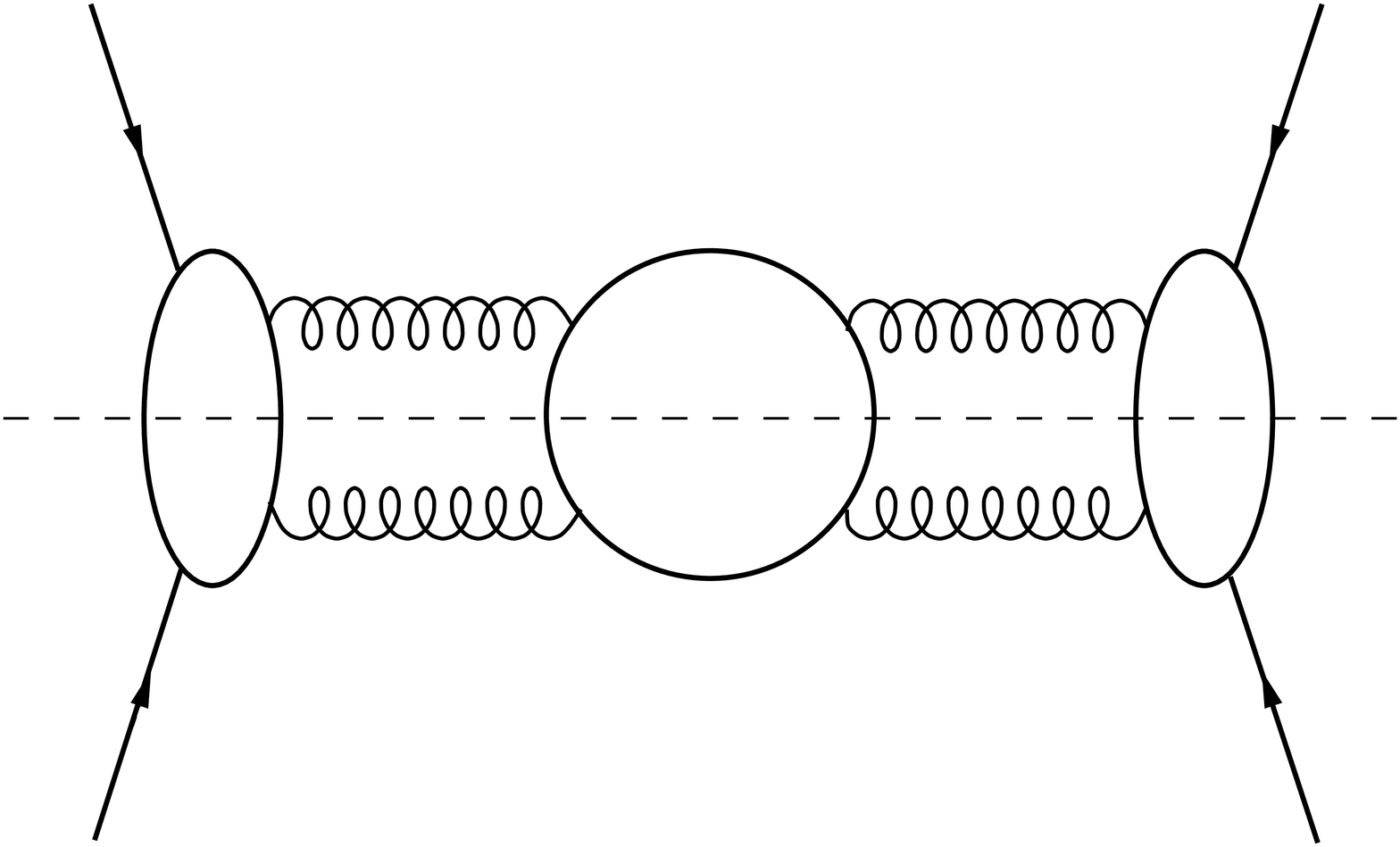}}
  \put(10,20){$h_{\pa}$}\put(7,1){\sf a}
  \put(66,20){$h_{\pb}$}\put(70,1){\sf b}
  \put(38,20){$\G_{\o}$}
  \put(1,8){$p_1$}\put(75,8){$p_2$}
  \put(22,12){$k_1$}\put(55,12){$k_2$}
\end{picture}
  \caption{\small\sl Double $\kk$-factorization of dijet differential
                     cross section.}
\label{hgh}
\end{figure}

Most of the NL$x$ calculations needed for the definition of the NL$x$
kernel have been performed \cite{fl89,cc97}
in the case of high-energy parton-parton
scattering, in which no physical hard scale is present. While the
total cross section of this process has severe power-like Coulomb
singularities, it is hoped that by fixing the transverse momenta
$\ku\,,\,\kd$ of the fragmentation jets (corresponding to the
virtualities of the exchanged gluons, Fig.~(\ref{hgh}), one is able to define
a two scale hard process. A similar but not identical procedure was
devised by Mueller and Navelet \cite{mn87} in hadron-hadron scattering
(cfr. Sec.~(\ref{cffp})).

\begin{figure}[h]
 \centering
\begin{picture}(156,45)
  \put(0,0){\includegraphics[width=156mm]{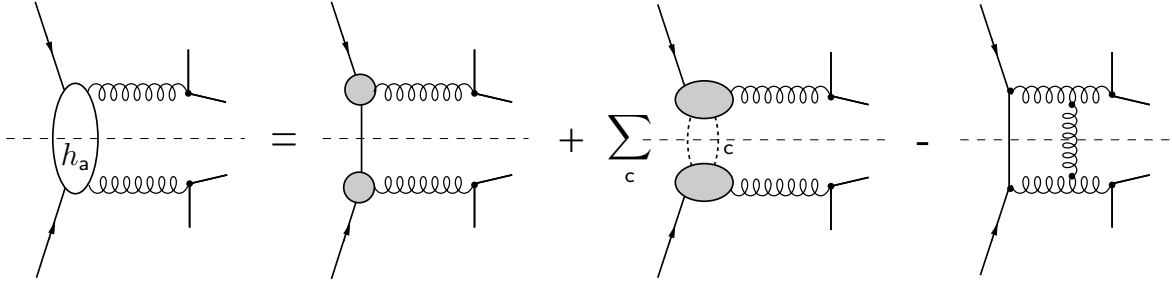}}
  \put(7.4,15.4){$h_{\pa}$}
  \put(80,18){$\ds\sum_{\sf c}$}
  \put(95.5,17.5){$_{\sf c}$}
 \end{picture}
  \caption{\small\sl Diagrammatic representation of the impact factor.
   In the sum over the possible final states, the outgoing parton
   index {\sf c} denotes the gluon {\sf g} for {\sf a = q} while it
   denotes the quark {\sf q} or the softer gluon for {\sf a = g}.}
\label{diagsottraz}
\end{figure}

We then consider high-energy scattering of
two partons {\sf a,b} ({\sf a,b}={\sf q}\,(quark), {\sf g}\,(gluon))
with momenta $p_1,p_2$ and we factorize it in impact factors and gluon
Green's function contributions as in Fig.~(\ref{hgh}), with the definition of
Fig.~(\ref{diagsottraz}).

By definition, the impact factors are free of high-energy gluon
exchanges, which are subtracted out as in Fig.~(\ref{diagsottraz}),
but still have collinear
singularities due to the initial massless partons which need to be
factored out. Therefore, the Regge-gluon exchanges are incorporated in
the Green's function $\G_{\o}$ of Fig.~(\ref{hgh}), and define
in a gauge-invariant way the
virtualities $\ku$ and $\kd$ which play the role of hard scales of the
process.

Formally, we can write \cite{c98} the colour averaged
differential cross section
\begin{equation}
 \frac{\dif\sigma_{\pa\pb}}{\du\,\dd}=%
 \int\frac{\dif\o}{2\pi i\o}\,h_{\pa}(\ku)\G_{\o}(\ku,\kd)
 h_{\pb}(\kd)\left(\frac{s}{s_0(\ku,\kd)}\right)^{\o}
 \label{fatt}
\end{equation}
where we have introduced, besides the $h$'s and $\G_{\o}$, also the
energy-scale $s_0(\ku,\kd)$ which is normalized by the condition
$s_0(\ku,\ku)=\ku^2$ and will be mostly chosen to be
$s_0=k_1 k_2$~($k_i=|\kk_i|$), for the reasons explained
below~\cite{fl98,cc98}. We adopt $\dk=\dif^{2+2\e}\kk/\pi^{1+\e}$ as
transverse space measure.
The validity of Eq.~(\ref{fatt}) at NL$x$ level has been so
far checked at one-loop \cite{c98} and two-loop \cite{cc98}
orders and needs be established at higher orders.

Let us start reviewing the leading $\log x$ (L$x$) results and the
NL$x$ ones of Refs.~\cite{c98} and \cite{cc98}.

We use the notation $p_1,p_2$ ($\l_1,\l_2$) for the initial parton's
momenta (helicities) and the indices $3,4$ (possibly $5$) for the
final ones, with the Sudakov parametrization 
\begin{align*}
  k_1&=p_1-p_3\;=\;z_1 p_1-\frac{\ku^2}{(1-z_1) s} p_2+k_{1\perp}\quad,\\
  k_2&=p_2-p_4\;=\;-\frac{\kd^2}{(1-z_2) s}p_1+z_2 p_2+k_{2\perp}\quad,
\end{align*}
where we have introduced Sudakov variables $z_i$ and
transverse spacelike vectors $k_{i\perp}$ perpendicular to the plane
$\langle p_1,p_2 \rangle:p_1\ps k_{i\perp}=0=p_2\ps k_{i\perp}$
with $D-2$ euclidean components $\kk_i:\kk_i^2=-k_{i\perp}^2>0$.
 
\begin{figure}[h]
 \centering
\begin{picture}(34,42)
  \put(0,0){\includegraphics[width=34mm]{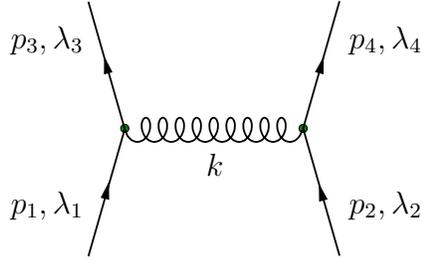}}
  \put(-10,6){$p_1,\l_1$}\put(35,6){$p_2,\l_2$}
  \put(-10,28){$p_3,\l_3$}\put(35,28){$p_4,\l_4$}
  \put(16,11){$k$}
\end{picture}
  \caption{\small\sl Born diagram of high energy scattering.}
\label{diagborn}
\end{figure}

At Born level, the cross section (\ref{fatt}) is just non-abelian
Coulomb scattering (Fig.~\ref{diagborn}),
corresponding to the high-energy amplitude 
\begin{equation}
 {\cal M}^{(0)}=\frac{2s}{t}\,g^2\,\t_{\pa}\ps \t_{\pb}\,
 \delta_{\l_3\l_1}\delta_{\l_4\l_2}\quad, \label{aborn}
\end{equation}
where helicity conserving factors are exhibited, but colour indices
are understood.

Upon helicity and colour averaging and by using the
$(4+2\e)$-dimensional phase space for two particle final states
\begin{equation}
 \dif\phi^{(2)}=\frac{\pi}{(4\pi)^{2+\e}}\,\frac{1}{s}\,\dk\quad,
\label{sfdue}
\end{equation}
Eq.~(\ref{aborn}) yields the differential cross section
\begin{align}
  \frac{\dif\sigma^{(0)}_{\pa\pb}}{\dk}
 &=\frac{\pi}{N_c^2-1}\,C_{\pa}C_{\pb}\,
  \left(\frac{2\as N_{\e}}{\kk^2\,\mu^{2\e}}\right)^2
  =h_{\pa}^{(0)}({\kk})\,h_{\pb}^{(0)}({\kk})\quad, \label{sborn}\\
 &\quad C_{\pq}=C_F=\frac{N_c^2-1}{2N_c}=\frac{4}{3}
  \quad;\quad C_{\pg}=C_A=N_c=3
  \quad;\quad N_{\e}=\frac{(4\pi)^{\e/2}}{\Gamma(1-\e)}\quad, \nonumber
\end{align}
in terms of the dimensionless coupling constant
\begin{equation*}
 \as=\frac{g^2\Gamma(1-\e)(\mu^2)^{\e}}{(4\pi)^{1+\e}}
\end{equation*}
and of the Born impact factors
\begin{equation}
 h_{\pa}^{(0)}({\kk})=\sqrt{\frac{\pi}{N_c^2-1}}\;
 \frac{2C_{\pa} \as N_{\e}}{\kk^2\,\mu^{2\e}}\quad,    \label{hzero}
\end{equation}
$\mu$ being the renormalization scale.

\begin{figure}[h]
 \centering
\begin{picture}(34,42)
  \put(0,0){\includegraphics[width=34mm]{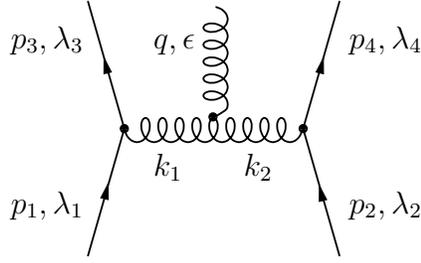}}
  \put(-10,6){$p_1,\l_1$}\put(35,6){$p_2,\l_2$}
  \put(-10,28){$p_3,\l_3$}\put(35,28){$p_4,\l_4$}
  \put(9,11){$k_1$}\put(21,11){$k_2$}
  \put(9,28){$q,\po$}
\end{picture}
  \caption{\small\sl Real gluon emission in central region.}
\label{qqg}
\end{figure}

At one-loop order, real emission and virtual contributions must be
added to the Born cross section. At L$x$ level,
the high-energy kinematics favours the emission of an additional gluon
(Fig.~\ref{qqg}) of momentum $p_5=q$ and polarization
$\po$ in the central region
$q^+\sim q^-\sim |\qq|$~. The corresponding amplitude is
\begin{equation}
 {\cal M}^{(L)}_{\pa\pg\pb}=g^3\,\frac{2s}{|\ku|^2|\kd|^2}\,
 \t_{\pa}^a \t_{\pb}^b\,
 \f^{abc}\,J\ps\po\,\delta_{\l_3\l_1}\delta_{\l_4\l_2}  \label{arl}
\end{equation}
where $q^{\mu}=k_1^{\mu}+k_2^{\mu}$ and
\begin{equation}
 J^{\mu}(k_1,k_2)=-k_1^{\mu}+k_2^{\mu}
 +\frac{p_1^{\mu}}{p_1\ps q}(\qq^2-\ku^2)
 -\frac{p_2^{\mu}}{p_2\ps q}(\qq^2-\kd^2) \label{verlip}
\end{equation}
is the Lipatov current \cite{bfkl77}.

After polarization and colour averaging, and by using the three-body
phase space
\begin{equation}
 \dif\phi^{(3)}=\frac{\pi}{(4\pi)^{4+2\e}}\,
 \frac{\dif z_1\,\du\,\dd}{s^2 z_1 (1-z_1)}\quad,    \label{sftre}
\end{equation}
we obtain the differential cross section
\begin{equation}
 \left.\frac{\dif\sigma^{(L)}_{\pa\pg\pb}}{\dif z_1\,\du\,\dd}
 \right)_{\rm real}=
 h^{(0)}_{\pa}(\ku)h^{(0)}_{\pb}(\kd)\,
 \frac{\ab}{\qq^2\Gamma(1-\e)\mu^{2\e}}\,\frac{1}{z_1}
 \quad;\quad\ab=\frac{\as N_c}{\pi}   \label{sezdom}
\end{equation}
which, upon the $z_1$ integration with infrared boundary $z_1>s_0/s$, becomes
\begin{equation}
 \left.\frac{\dif\sigma^{(L)}_{\pa\pg\pb}}{\du\dd}\right)_{\rm real}=
 h^{(0)}_{\pa}(\ku)h^{(0)}_{\pb}(\kd)\,
 \frac{\ab}{\qq^2\Gamma(1-\e)\mu^{2\e}}\,\log\frac{s}{s_0}\quad.\label{srl}
\end{equation}

At this stage, the scale $s_0$ is totally undetermined (pure phase
space would yield $\frac{\qq^2}{s}<z_1<1-\frac{\ku^2}{s}$)
because the NL$x$ constant has not been calculated yet (cfr.~Sec.{\ref{ifnla}).

The L$x$ virtual one-loop corrections give rise to reggeization of
the exchanged gluons, so that their propagator is modified by a factor
\cite{ffqk96}
\begin{equation}
 \left(\frac{s}{\kk^2}\right)^{\o(\kk^2)}=1+
 \o^{(1)}(\kk^2)\log\frac{s}{\kk^2}+{\cal O}(\as^2)
\label{regge}
\end{equation}
where
\begin{equation}
 \o^{(1)}(\kk^2)=-\frac{g^2 C_A}{(4\pi)^{2+\e}}\,\kk^2
 \int\frac{\dif[{\boldsymbol p}]}{{\boldsymbol p}^2(\kk-{\boldsymbol p})^2}=
 -\frac{\ab}{2\e}\frac{\Gamma^2(1+\e)}{\Gamma(1+2\e)}
 \left(\frac{\kk^2}{\mu^2}\right)^{\e}\quad.
 \label{omega}
\end{equation}
Therefore, the one-loop contribution to the elastic cross-section becomes
\begin{equation}
 \left.\frac{\dif\sigma^{(L)}_{\pa\pb}}{\dk}\right)_{\rm virtual}=
 \frac{\dif\sigma^{(0)}_{\pa\pb}}{\dk}\,2\o^{(1)}(\kk^2)
 \log\frac{s}{\kk^2}  \label{svl}
\end{equation}
and the total leading contribution, up to one-loop order, becomes
\begin{equation}
 \frac{\dif\sigma^{(L)}_{\pa\pb}}{\du\,\dd}=
 h^{(0)}_{\pa}(\ku)h^{(0)}_{\pb}(\kd)
 \left[\dq+\ab K_0(\ku,\kd)\log\frac{s}{s_0}\right]\label{sl}
\end{equation}
where
\begin{equation}
 \ab K_0(\ku,\kd)=\frac{\ab}{\qq^2\Gamma(1-\e)\mu^{2\e}}
 +2\o^{(1)}(\ku^2)\,\dq\quad;\quad
 \dq=\pi^{1+\e}\delta^{2+2\e}(\qq)  \label{nucleo}
\end{equation}
is the leading BFKL kernel \cite{bfkl77}.

The result (\ref{sl}) checks with the representation (\ref{fatt}) by
defining the Green's function
\begin{equation}
  \G_{\o}^{(L)}=\left(1-\frac{\ab}{\o}K_0\right)^{-1}   \label{green}
\end{equation}
which satisfies the L$x$ BFKL equation.
Next, we shall consider NL$x$ corrections.


\section{Impact factors at NL$x$ accuracy}\label{ifnla}

At one-loop NL$x$ level, we should calculate the constant contributions
to the differential cross section (\ref{fatt}), which require
an accurate treatment
of the fragmentation region. Constant terms arise from both the one-loop
corrections to the impact factors and possible constant kernels in the gluon
Green's function. In fact, the general NL$x$ form of $\G_{\o}$ is \cite{cc98}
\begin{equation}
 \G_{\o}=(1+\ab H_L)\left[1-\frac{\ab}{\o}(K_0+K_{NL})\right]^{-1}\!(1+\ab H_R)
 \label{scomp}
\end{equation}
where we have introduced the operator factors $H_L , H_R$ and the NL$x$ kernel
\begin{equation}
 K_{NL}=-b\,\as\log\frac{\ku^2}{\mu^2}\,K_0+\as K_1
 \quad;\quad b=\frac{11N_c-2n_f}{12\pi}     \label{knl}
\end{equation}
which consists in the running coupling contribution, proportional to
$K_0$ with a $\beta$-function coefficient, and in the renormalization
scale independent part $\as K_1$.

The one-loop expansion of Eq.~(\ref{fatt}), taking into account the definition
(\ref{scomp}), reads
\begin{align}
 &h_{\pa}^{(0)}(\ku)\;\ab K_0(\ku,\kd)\;h_{\pb}^{(0)}(\kd)\,
  \log\frac{s}{s_0(\ku,\kd)}+     \nonumber \\
 &+\ab\Big(h_{\pa}^{(1)}(\ku)\dq+h_{\pa}^{(0)}(\ku)H_L(\ku,\kd)\Big)
  h_{\pb}^{(0)}(\kd)\;+\;\mbox{\sf a}\leftrightarrow\mbox{\sf b}\quad.
 \label{svil}
\end{align}
We see, therefore, that the identification of the one-loop impact
factors $h_{\pa}^{(1)},h_{\pb}^{(1)}$ is affected by a factorization
scheme ambiguity, due to both the choice of the scale $s_0$ and of the
kernel $H_L$ ($H_R$).

In Ref.~\cite{c98} we have defined the impact
factors by the following factorization procedure:
\begin{itemize}
 \item[($i$)] Subtract the leading term with a reference scale,
  consistent with the infrared
  and collinear properties of the process, for which we can set
  $H_L=H_R=0$;
 \item[($ii$)] Interpret the remaining constant at
  single-$\kk$~factorization level as the one-loop correction $h^{(1)}$.
\end{itemize}

In order to perform the subtraction ($i$) it is convenient to define
the fragmentation 
vertex\footnote{In this paper $F_{\pa}$ differs from
                that of Ref.~\cite{cc98} by a normalization factor
                $\as/z_1\qq^2$.}
$F_{\pa}(z_1,\ku,\kd)$ which contains all (real and virtual)
one-loop squared matrix elements, after factorization of the exchanged
Regge-gluon $\kd$. In the real emission contribution to $F_{\pa}$, we
fix the momentum fraction $z_1$ and the transverse momentum $\qq$ of
the emitted parton {\sf c}. In order to avoid double counting we also
set {\sf c = g} for {\sf a = q} and {\sf c = q} or {\sf c} = the
softer gluon in the case {\sf a = g}.
Then in Ref.~\cite{c98} the leading term was
subtracted out as follows (here and in the sequel $q=|\qq|,k_i=|\kk_i|$):
\begin{equation}
 F_{\pa}(z_1,\ku,\kd)-\ab h_{\pa}^{(0)}(\ku)\,
 \frac{\Theta(q-z_1 k_1)}{z_1\qq^2\Gamma(1-\e)\mu^{2\e}}
 =\ab F_{\pa}^{(1)}(z_1,\ku,\kd)        \label{fauno}
\end{equation}
and the impact factor contribution was defined by $z_1$ and $\ku$
integration:
\begin{equation}
 h_{\pa}^{(1)}(\kd)=\int\du\int_0^1\dif z_1\,F_{\pa}^{(1)}(z_1,\ku,\kd)
 \quad.   \label{huno}
\end{equation}

Note that the theta-function in the subtraction term in
Eq.~(\ref{fauno}) cuts off the leading contribution $1/z_1\qq^2$ for
$q<k_1$, by the angular ordering constraint $z_1<q/k_1$.
This insures that $F_{\pa}^{(1)}$ be both integrable at $z_1=0$ and
without spurious $\qq=0$ singularities.

In fact, the fragmentation function $F_{\pa}$ is expected to have the
leading $1/z_1$ singularity for $z_1\rightarrow 0$, but no $\qq^2=0$
singularity\footnote{The $1/\qq^2$ terms, pertaining to the Sudakov
                   fragmentation region, actually cancel out with virtual
                 corrections in Eq.~(\ref{huno}) (App.~\ref{integrali}).}.
This coherence effect, first noticed by one of us \cite{c88}, is to be
inserted in the subtraction term also, in order to avoid spurious
singularities in Eq.~(\ref{huno}).

With the definition (\ref{fauno}), the impact factor (\ref{huno}) is
expected to have the $\ku^2=0$ collinear singularities of $F_{\pa}$
only, except the ones subtracted out with the leading term.
In the quark case, the left over collinear singularity is the one
provided by the nonsingular part of the zero-moment $\tilde{P}_{gq}(\o=0)$
(see Sec.~(\ref{cffp})).

By integrating Eq.~(\ref{fauno}) in the fragmentation region of
{\sf a} ($\frac{q}{\sqrt{s}}<z_1<1$) we obtain
\begin{equation}
 \int_{q/\sqrt{s}}^1\dif z_1\,F_{\pa}(z_1,\ku,\kd)=
 \ab\frac{h_{\pa}^{(0)}(\ku)}{\qq^2\Gamma(1-\e)\mu^{2\e}}\left[\log
 \frac{\sqrt{s}}{{\rm Max}(q,k_1)}+h_{\pa}(\qq,\ku)\right]
 \label{intfa}
\end{equation}
where $h_{\pa}$, which is characterized by its vanishing at $\qq=0$
\cite{c98}, is related to the proper anomalous dimension for $\ku=0$,
as shown in more explicit form in Sec.~(\ref{qqcoll}). Therefore, the
definition (\ref{huno}) corresponds to setting $H_L=H_R=0$ for the
reference scale
\begin{equation}
 s_>={\rm Max}(q,k_1){\rm Max}(q,k_2) \label{scalamax}
\end{equation}
which takes into account the coherence effect mentioned before.

A similar definition of the impact factors can be adopted for
colourless sources also, as in double-DIS or quarkonium processes. In
this case, there is an additional dependence on the $\o$ variable and
on the external hard scale(s) so that, at leading level, the impact
factors 
\begin{equation}
 h_{\pa}^{(0)}=h_{\pa,\o}^{(0)}(Q_1,\ku)=\frac{1}{Q_1^2}\,f^{\pa}_{\o}
 \left(\frac{\ku^2}{Q_1^2}\right)  \label{hbianco}
\end{equation}
yield a nontrivial $\ku$-dependence, which has been explicitly
computed in several processes \cite{cc96a}. In Eq.~(\ref{hbianco})
the function $f_{\pa}$ has the role of setting $k_1$ of order $Q_1$ in
the total cross section, and has no $\ku^2=0$ singularity at all,
because there is no initial colour 
charge\footnote{We are indebted to V. Fadin for a discussion on this point.}.

Therefore, by translating angular ordering into the $\o$-space
threshold factor $(q/k_1)^{\o}$, the analogue of Eq.~(\ref{fauno}) becomes
\begin{equation}
 F_{\pa,\o}(Q_1,\ku,\kd)-\ab h_{\pa,\o}^{(0)}(Q_1,\ku)\,
 \frac{1}{\o\qq^2\mu^{2\e}}\left(\frac{q}{k_1}\right)^{\o\,\Theta_{k_1q}}
 =\ab F_{\pa,\o}^{(1)}(Q_1,\ku,\kd)
 \label{fadue}
\end{equation}
with the definition
\begin{equation}
 h_{\pa,\o}^{(1)}(Q_1,\kd)=\int\du\int_0^1\dif
 z_1\,F_{\pa,\o}^{(1)}(Q_1,z_1,\ku,\kd)\quad.   \label{hdue}
\end{equation}

In this case, no $\qq=0$ nor $\ku=0$ singularities are expected in
$F_{\pa}$, so that it is again important that the leading term be
subtracted out with the angular ordering constraint.

We thus conclude that the definition of the
impact factors in Eqs.~(\ref{huno}) and (\ref{hdue}),
on the basis of the subtraction in
Eqs.~(\ref{fauno}) and (\ref{fadue}) defines a self-consistent
$\kk$-factorization scheme of Eq.~(\ref{fatt}) for both coloured and
colourless sources.

The question then arises: since this factorization scheme singles
out the scale $s_>$ of Eq.~(\ref{scalamax}), does this mean that
$H_L=H_R=0$ in general?

The answer to the question above depends on the identification of the
scale $s_0$, possibly different from $s_>\,$, for which the
representation (\ref{fatt}) is assumed to be valid to all orders. Such
all order identification has not been rigorously provided yet. However
the scale $s_>$ has the defect of not being factorized in its
$\ku,\kd$ dependence. This fact is argued to contradict multi-Regge
factorization of production amplitudes \cite{bw70}, which implies
short-range correlations of the $\kk$'s. Furthermore, the use of
$O(2,1)$ variables \cite{cdm69} implies that Regge behaviour in the
energy $s$ should involve the boost
\begin{equation}
 \cosh(\zeta)=\frac{s}{2k_1k_2}\quad,  \label{spinta}
\end{equation}
thus suggesting the choice $s_0=k_1k_2$, adopted in Refs.~\cite{fl98}
and \cite{cc98}.

If $s_0=k_1k_2$ is assumed, then Eq.~(\ref{intfa}) implies that
$H_R=H_L^{\dagger}=H$ is nonvanishing, and is given by the expression
\cite{cc98}
\begin{equation}
 H(\ku,\kd)=-\frac{1}{\qq^2\mu^{2\e}}\,\log\frac{q}{k_1}\;
 \Theta_{qk_1}\qquad,\qquad(s_0=k_1k_2)\quad. \label{acca}
\end{equation}

By replacing Eq.~(\ref{acca}) in the form (\ref{scomp}) of the gluon
Green's function, the representation (\ref{fatt}) can be expanded at
two-loop level in order to derive the NL$x$ kernel $K_1$ \cite{cc98} for
the scale $s_0=k_1k_2$, from known two-loop calculations \cite{cc97}.
The scale changes from  $s_0=k_1k_2$ to other factorized combinations,
like $s_0=k_1^2$ ($s_0=k_2^2$), which are relevant for
$k_1^2\gg k_2^2$ ($k_1^2\ll k^2_2$), were described in
Ref.~\cite{cc98}.

An important point concerns the possible choice of different
$\kk$-factorization schemes. In fact the precise form of the subtraction
term in Eqs.~(\ref{fauno}) and (\ref{fadue}) is not uniquely
determined, away from $z_1=0$. Alternative forms exist, which reduce
to the leading term for $z_1\rightarrow0$, and vanish for
$q\rightarrow0$. Switching to such alternative forms defines an
alternative $\kk$-factorization scheme which differs from the above by a
universal and finite impact factor renormalization. For instance, one
can switch \cite{c98} from the reference
scale $s_>$ to the scale ${\rm Max}(\ku^2,\kd^2)$ by the change in the
constant piece $h_{\pa}$ in Eq.~(\ref{intfa})
\begin{equation}
 \delta h(\qq,\ku)=\log\frac{k_2}{k_1}\;\Theta_{k_2 k_1}-
 \log\frac{q}{k_1}\;\Theta_{q k_1}    \label{deltah}
\end{equation}
which, upon $\ku$-integration, implies an {\sf a}-independent change in
the impact factors
\begin{equation}
 \delta h_{\pa}^{(1)}=\ab\,\frac{1}{2}\,\psi^{\prime}(1)
 \,h_{\pa}^{(0)}\quad.      \label{dhuno}
\end{equation}
Correspondingly, the gluon Green's function (\ref{scomp}) changes
in the constant kernels by the quantities
\begin{equation}
 \delta H_L(\ku,\kd)=-\delta h(\qq,\ku)\qquad,\qquad
 \delta H_R(\ku,\kd)=-\delta h(\qq,\kd)  \label{dacca}
\end{equation}
so as to keep the overall cross section unchanged.

In the following, we apply the $\kk$-factorization scheme defined in
Eqs.~(\ref{fauno}) and (\ref{huno}) to the calculation of the one-loop
corrections to the partonic impact factors.


\section{Quark-parton collision}\label{qqcoll}

\begin{figure}[h]
 \centering
\begin{picture}(34,41)
  \put(0,0){\includegraphics[width=34mm]{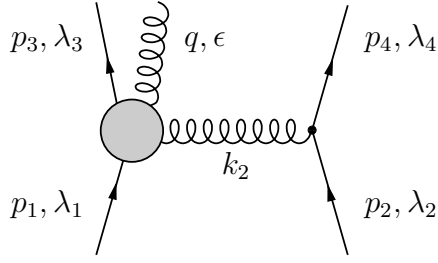}}
  \put(-11,6){$p_1,\l_1$}\put(36,6){$p_2,\l_2$}
  \put(-11,28){$p_3,\l_3$}\put(36,28){$p_4,\l_4$}
  \put(17,11){$k_2$}
  \put(12,28){$q,\po$}
\end{picture}
  \caption{\small\sl Real gluon emission amplitude in quark {\sf a}
                     fragmentation region.}
\end{figure}

The calculation for the quark-parton scattering was done in
Ref.~\cite{c98} and here we sketch the main results. The real emission
amplitude, assuming the gluon in the quark {\sf a} fragmentation
region and adopting the Feynman gauge, is
\begin{equation}
 \M_{\pq\pg\pb}=\poc\,^{\mu}\,\overline{u}_3 A_{\mu\nu}^{ab}u_1
 \,\frac{1}{k_2^2}\;
 g\,2p_2^{\nu}\,\t_{\pb}^b\,\delta_{\l_2\l_4} \label{aqgq}
\end{equation}
where $A_{\mu\nu}^{ab}$ is diagrammatically represented in
Fig.~(\ref{tensoregq}).
\begin{figure}[h]
 \centering
\begin{picture}(140,47)
  \put(0,0){\includegraphics[width=140mm]{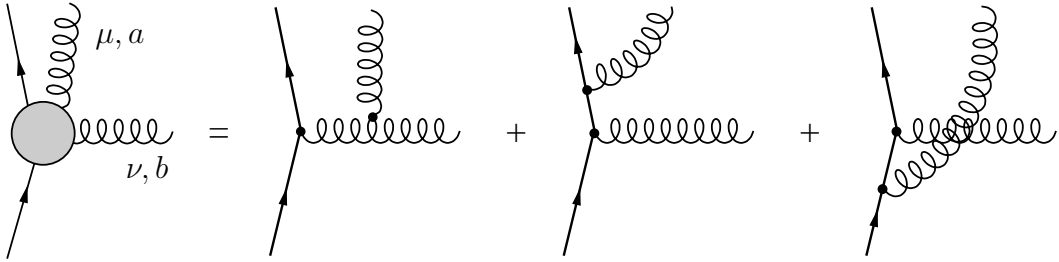}}
  \put(16,11){$\nu,b$}\put(12,29){$\mu,a$}
\end{picture}
  \caption{\small\sl Quark to gluon fragmentation tensor $A_{\mu\nu}^{ab}$.}
\label{tensoregq}
\end{figure}

After a simple calculation,
the differential cross section in this kinematical region results to be
\begin{align}
 \frac{\dif\sigma_{\pq\pg\pb}}{\dif z_1\,\du\,\dd}=\;&
 h_{\pq}^{(0)}(\ku)\,h_{\pb}^{(0)}(\kd)\,
 \frac{\P_{\pg\pq}(z_1,\e)}{\Gamma(1-\e)\mu^{2\e}}\times\nonumber\\
 &\times\left(\frac{C_A\as}{\pi}\,
 \frac{(1-z_1)\qq\ps(\qq-z_1\kd)}{\qq^2\,(\qq-z_1\kd)^2}+
 \frac{C_F\as}{\pi}\,\frac{z_1^2\ku^2}{\qq^2\,(\qq-z_1\kd)^2}\right)
 \label{sqgq}
\end{align}
where
\begin{equation}
 \P_{\pg\pq}(z_1,\e)=\frac{1}{2z_1}\left[1+(1-z_1)^2+\e z_1^2\right]
 \label{pgq}
\end{equation}
is related to the quark to gluon splitting function. We can easily check
that $\dif\sigma_{\pq\pg\pb}$ matches
$\dif\sigma^{(L)}_{\pq\pg\pb}$ (\ref{sezdom})
in the central region $z_1\ll 1$,
so that we can take it as the right differential cross section in the
whole positive $q^{\mu}$ rapidity range
\begin{equation}
 y_5=\log\frac{z_1\sqrt{s}}{q}\geqslant0\quad\iff\quad
 z_1\geqslant\frac{q}{\sqrt{s}}\quad.       \label{rapidita}
\end{equation}
In the remaining half phase-space $y_5<0$ the cross section has the
same expression provided we exchange
$\ku\leftrightarrow\kd\;,\;z_1\leftrightarrow z_2$, where $z_2$ is the
momentum fraction of $q$ with respect to $p_2$.

It is worth noting that the splitting function (\ref{pgq}) is factored
out in Eq.~(\ref{sqgq}) even outside the collinear regions
$\qq^2\ll\ku^2\simeq\kd^2$ and $\ku^2\ll\qq^2\simeq\kd^2$ thus
suggesting a smooth extrapolation between collinear and Regge regions
\cite{c88,cfm90}.

The correction to the cross section due to virtual emission, including
subleading effects, for general parton-parton scattering, can be
extracted from the amplitude of Ref.~\cite{ffqk96} 
\begin{equation}
 \M_{\pa\pb}=2s\Big(g\,\t_{\pa}^c\delta_{\l_3\l_1}\Big)
 \Big(1+\Gamma^{(+)}_{\pa\pa}\Big)\frac{1}{t}
 \left[1+\o(-t)\log\frac{s}{-t}\right]\Big(1+\Gamma^{(+)}_{\pb\pb}\Big)
 \Big(g\,\t_{\pb}^c\delta_{\l_4\l_2}\Big)  \label{corrvirt}
\end{equation}
where, in the quark case,
\begin{equation}
 \Gamma^{(+)}_{\pq\pq}=\o^{(1)}(\ku^2)\left[
 \frac{11}{12}-\left(\frac{85}{36}+\frac{\pi^2}{4}\right)\e-
 \frac{n_f}{N_c}\left(\frac{1}{6}-\frac{5}{18}\e\right)+
 \frac{C_F}{N_c}\left(\frac{1}{\e}-\frac{3}{2}+4\e\right)\right]
 \label{gammaqq}
\end{equation}
is the quark-quark-reggeon vertex correction in the helicity
conserving channel. Therefore, we have to add the virtual term
$h^{(0)}_{\pq}(\ku)\,2\Gamma^{(+)}_{\pq\pq}\delta(1-z_1)\dq$
to the fragmentation vertex $F_{\pa}(z_1,\ku,\kd)$.

The $C_F$ term of real emission is not
singular at $z_1=0$ and doesn't contribute to the logarithmic part of
the cross section; therefore, it can be immediately integrated over
$z_1$ and $\ku$ at fixed $\kd$. Since the result exactly cancels the
$C_F$ term of virtual emission, we can write the total
fragmentation vertex for the $C_A$ term only:
\begin{align}
 \hspace{-9mm}F_{\pq}(z_1,\ku,\kd)=h^{(0)}_{\pq}(\ku)&\left\{\dq
  \delta(1-z_1)2\o^{(1)}(\ku^2)\left[
  \frac{11}{12}-\left(\frac{85}{36}+\frac{\pi^2}{4}\right)\!\e-
  \frac{n_f}{N_c}\left(\frac{1}{6}-\frac{5}{18}\e
  \right)\right]\right.   \nonumber \\
 &+\left.\frac{\P_{\pg\pq}(z_1,\e)}{\Gamma(1-\e)\mu^{2\e}}\,\frac{\ab}{\qq^2}
 \,\frac{(1-z_1)\qq\ps(\qq-z_1\kd)}{(\qq-z_1\kd)^2}\right\}\quad.
 \label{effeq}
\end{align}

We can see how the features anticipated in Eqs.~(\ref{fauno}) and
(\ref{huno}) are realized by the explicit expression (\ref{effeq}). In
fact, $F_{\pq}$ has no $\qq^2$ singularity, except at $z_1=0$, where it
reduces to the leading term to be subtracted out.

By using the procedure of Eqs.~(\ref{fauno})-(\ref{intfa}) we arrive
at the definition
\begin{equation}
 \frac{h_{\pq}^{(0)}(\kd)}{\qq^2\Gamma(1-\e)\mu^{2\e}}\;h_{\pq}(\qq,\ku)
 =\int_0^1\dif z_1\,F_{\pq}^{(1)}(z_1,\ku,\kd)
\end{equation}
in which $h_{\pq}(\qq,\ku)$ vanishes at $\qq=0$.
By integration at fixed $\kd$, we obtain (App.~\ref{integrali}) the
one-loop correction to the quark impact factor, as follows:
\begin{align}
 h_{\pq}^{(1)}(\kd)&=h_{\pq}^{(0)}(\kd)\,\o^{(1)}(\kd^2)\left[\left(
 \frac{11}{6}-\frac{n_f}{3N_c}\right)+\left(\frac{3}{2}-\frac{1}{2}\e\right)
 -\frac{\K}{N_c}\e\right]\quad,   \label{hunoq}\\
 \K&=\left[N_c\left(\frac{67}{18}-\frac{\pi^2}{6}\right)-\frac{5n_f}{9}
  \right]\quad. \label{k}
\end{align}


\section{Gluon-parton collision}\label{ggcoll}

We consider now gluon-parton scattering. In this case we have to
distinguish two different final states for three-particle production.
In addition to the parton {\sf b} there may be:
\begin{itemize}
 \item[{\it i})] two gluons;
 \item[{\it ii})] a quark-antiquark pair.
\end{itemize}

\subsection{{\sf gb $\rightarrow$ ggb}}

\begin{figure}[h]
 \centering
\begin{picture}(34,40)
  \put(0,0){\includegraphics[width=34mm]{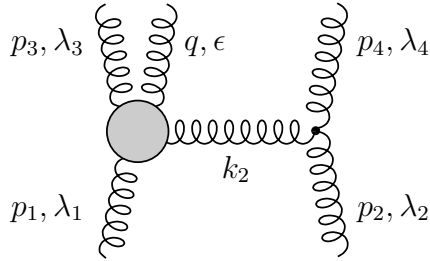}}
  \put(-11,6){$p_1,\l_1$}\put(35,6){$p_2,\l_2$}
  \put(-11,28){$p_3,\l_3$}\put(35,28){$p_4,\l_4$}
  \put(17,11){$k_2$}
  \put(12,28){$q,\po$}
\end{picture}
  \caption{\small\sl Real gluon emission amplitude in gluon {\sf a}
                     fragmentation region.}
\label{diagggg}
\end{figure}

Assume that the additional gluon is emitted in the gluon {\sf a}
fragmentation region. The scattering amplitude in the Feynman gauge
is given by
\begin{equation}
 \M_{\pg\pg\pb}=
 \po_1^{\mu_1}\,\poc\,_3^{\mu_3}\,\poc\,_5^{\mu_5}
 \;A_{\mu_1\mu_3\mu_5\,\nu}^{a_1a_3a_5\,b}\,\frac{1}{k_2^2}\;
 g\,2p_2^{\nu}\,\t_{\pb}^b\,\delta_{\l_4\l_2} \label{aggg}
\end{equation}
where the amplitude $A$, corresponding to the diagrams of
Fig.~(\ref{tensoregg}) was found in Ref.~\cite{fl89} to be
\begin{align}
 A_{\mu_1\mu_3\mu_5\,\nu}^{a_1a_3a_5\,b}\,2p_2^{\nu}&=
 4g^2\,g_{\mu_1\mu_3}\Big[\f^{a_1a_5e}\f^{ea_3b}D_{\mu_5}(-p_1,p_3,p_5)+
 \f^{a_3a_5e}\f^{ea_1b}D_{\mu_5}(p_3,-p_1,p_5)\Big]+\nonumber \\
 &\qquad+\,\begin{picture}(23,9)(0,5)
               \put(0,4.5){$\Bigg($}\put(19,4.5){$\Bigg)$}
               \put(4,10){$p_3\leftrightarrow p_5$}
               \put(3.7,5){$\mu_3\leftrightarrow\mu_5$}
               \put(4,0){$a_3\leftrightarrow a_5$}
           \end{picture}
        +\,\begin{picture}(23,9)(0,5)
               \put(0,4.5){$\Bigg($}\put(21,4.5){$\Bigg)$}
               \put(3,10){$-p_1\leftrightarrow p_5$}
               \put(6,5){$\mu_1\leftrightarrow\mu_5$}
               \put(6.3,0){$a_1\leftrightarrow a_5$}
           \end{picture} \quad, \label{amumu}
\end{align}
where the current
\begin{equation}
 D^{\mu}(x,y,z)=\frac{1}{x\ps y}\left[\left(y\ps z-p_2\ps p_4\,
 \frac{p_2\ps y}{p_2\ps z}\right)p_2^{\mu}+\frac{p_2\ps y}{x\ps z}
 (y\ps z-p_2\ps p_4)x^{\mu}+(x\ps p_2)y^{\mu}\right]     \label{d}
\end{equation}
is a function of the momenta $x^{\alpha},y^{\alpha},z^{\alpha}$, with
$x\ps y=x^{\alpha}y_{\alpha}$.
\begin{figure}[h]
 \centering
  \caption{\small\sl Gluon to gluon fragmentation tensor 
           $A_{\mu_1\mu_3\mu_5\nu}^{a_1a_3a_5b}$.}
\begin{picture}(140,47)
  \put(1,0){\includegraphics[width=139mm]{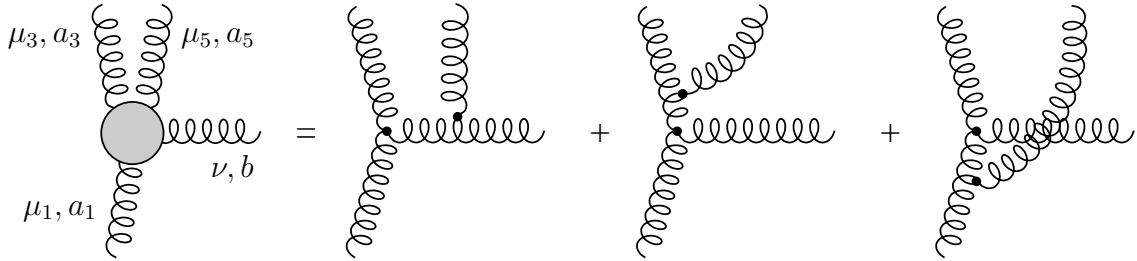}}
  \put(17,11){$\nu,b$}\put(13,29){$\mu_5,a_5$}
  \put(-8,6){$\mu_1,a_1$}\put(-10,29){$\mu_3,a_3$}
\end{picture}
\label{tensoregg}
\end{figure}

The squared helicity amplitudes corresponding to Eq.~(\ref{amumu})
were given in Ref.~\cite{ds97}. Since we work in $D=4+2\e$ dimensions,
we perform explicitly the polarization sum on Eq.~(\ref{aggg}) in
App.~\ref{integrali}. The averaged squared matrix element that
we obtain is actually independent of the spacetime dimensionality
(i.e. $\e$-independent) so that we finally obtain the same formula as
in Ref.~\cite{c98}:
\begin{align}
 \frac{\dif\sigma_{\pg\pg\pb}}{\dif z_1\,\du\,\dd}=\;&
 h_{\pg}^{(0)}(\ku)\,h_{\pb}^{(0)}(\kd)\,
 \frac{\P_{\pg\pg}(z_1)}{\Gamma(1-\e)\mu^{2\e}}\times\nonumber\\
 &\times\frac{C_A\as}{\pi}\,
 \frac{z_1^2\ku^2+(1-z_1)^2\qq^2+z_1(1-z_1)\ku\ps\qq}{%
 \qq^2\;(z_1\ku+(1-z_1)\qq)^2}\quad.    \label{sggg}
\end{align}
This expression explicitly exhibits the symmetry in the exchange
$-\ku\leftrightarrow\qq\;;\;z_1\leftrightarrow 1-z_1$ due to the identity
of the gluons emitted in the fragmentation region of gluon {\sf a},
but will be used for the softer gluon only $(z_1<\frac{1}{2})$, in
order to avoid double counting.

The function
\begin{equation}
 \P_{\pg\pg}(z_1)=\P_{\pg\pg}(1-z_1)=\frac{1+z_1^4+(1-z_1)^4}{2z_1(1-z_1)}
 \label{pgg}
\end{equation}
is related to the gluon to gluon splitting function, and is factored out
in Eq.~(\ref{sggg}) in the whole fragmentation region,
as for quark scattering. In this case also the cross section matches
the leading expression (\ref{sezdom})
in the central region $z_1\ll1$, and thus is taken to be valid in
the half phase space (\ref{rapidita}) in which two of the three gluons
have positive rapidity.

\subsection{{\sf gb $\rightarrow$ q$\overline{\sf q}$b}}

The main contribution to the cross section from this kind of final
states is reached when the fermion pair belongs to the same
fragmentation vertex, as in Fig.~(\ref{diagqqg}.a).
Other  graphs like Fig.~(\ref{diagqqg}.b) in the case $\sf b=g$
are suppressed by a factor $|t|/s$ because of fermionic exchange.
\begin{figure}[h]
 \centering
\begin{picture}(121,48)(0,-10)
  \put(0,0){\includegraphics[width=121mm]{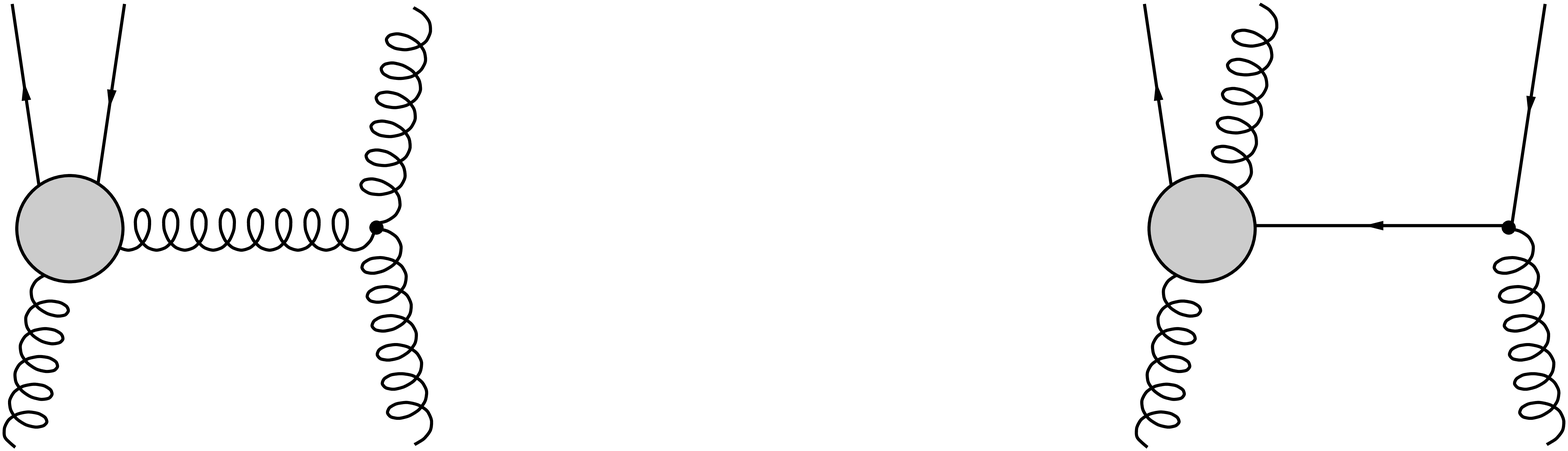}}
  \put(-11,6){$p_1,\l_1$}\put(35,6){$p_2,\l_2$}
  \put(-11,28){$p_3,\l_3$}\put(35,28){$p_4,\l_4$}
  \put(16,11){$k_2$}\put(11,28){$q,\po$}
  \put(15,-9){\sl a)}\put(102,-9){\sl b)}
\end{picture}
\label{diagqqg}
  \caption{\small\sl Quark antiquark emission amplitude for $\sf b=g$.
   a) The pair is emitted in quark {\sf a} fragmentation region;
   b) This kind of diagram is suppressed by a factor $|t|/s$.}
\end{figure}

Assuming {\sf q}$\overline{\sf q}$ being emitted in the fragmentation region
of gluon {\sf a}, and labelling {\sf q} with ``3'' and $\overline{\sf
q}$ with ``5'' ($p_5=q$), the corresponding amplitude is
\begin{equation}
 \M_{\pq\overline{\pq}\pb}=\po^{\mu}\,
 \overline{u}_3 B_{\mu\nu}^{ab}v_5\,\frac{1}{k_2^2}\;g\,2p_2^{\mu}\,
 \t_{\pb}^b\,\delta_{\l_4\l_2} \label{aqqg}
\end{equation}
where $B_{\mu\nu}^{ab}$ is the sum of the diagrams depicted in
Fig.~({\ref{tensoreqg}).
\begin{figure}[h]
 \centering
\begin{picture}(134,47)
  \put(0,0){\includegraphics[width=134mm]{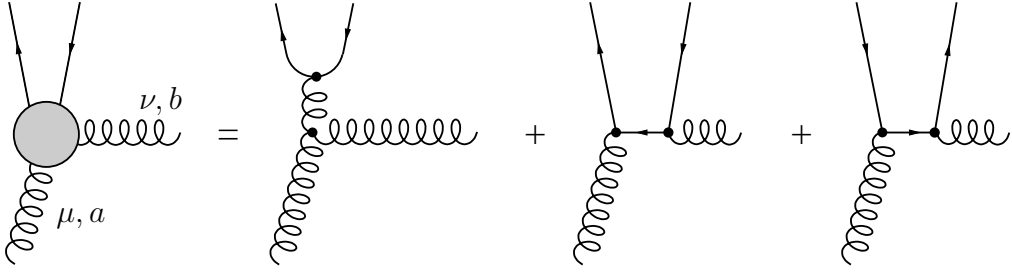}}
  \put(7,6){$\mu,a$}\put(18,21){$\nu,b$}
\end{picture}
  \caption{\small\sl Gluon to quark fragmentation tensor $B_{\mu\nu}^{ab}$.}
\label{tensoreqg}
\end{figure}

This calculation is very similar to the
{\sf qb $\rightarrow$ qgb} cross section, and yields
(App.~\ref{elementi})
\begin{align}
  \frac{\dif\sigma_{\pq\overline{\pq}\pb}}{\dif z_1\,\du\,\dd}
 &=n_f\,h_{\pb}^{(0)}(\kd)\,\frac{\as N_{\e}}{\ku^2\mu^{2\e}}\,
  \frac{\P_{\pq\pg}(z_1,\e)}{2\Gamma(1-\e)\mu^{2\e}}\times\nonumber\\
 &\qquad\times\left[\frac{C_A\as}{\pi}\,
  \frac{-z_1(1-z_1)\ku\ps\qq}{\qq^2\,(z_1\ku+(1-z_1)\qq)^2}
  +\frac{C_F\as}{\pi}\,\frac{1}{\qq^2}\right] \label{sqqg}
\end{align}
in which the function
\begin{equation}
 \P_{\pq\pg}(z_1,\e)=1-\frac{2z_1(1-z_1)}{1+\e}\quad, \label{pqg}
\end{equation}
related to the gluon to quark splitting function, is regular in
$z_1\in[0,1]$. Hence these final states do not produce any $\log s$
term, and need no subtraction.

The virtual correction to the amplitude, including NL$x$ effects, is
given by an expression analogue to Eq.~(\ref{corrvirt}) where
$\Gamma^{(+)}_{\pa\pa}$ is replaced by the gluon-gluon-reggeon vertex
correction~\cite{ffqk96} in the helicity conserving channel
\begin{equation}
 \Gamma^{(+)}_{\pg\pg}=\o^{(1)}(\ku^2)\left[
 \frac{1}{\e}-\frac{11}{12}+\left(\frac{67}{36}-\frac{\pi^2}{4}\right)\e
 +\frac{n_f}{N_c}\left(\frac{1}{6}-\frac{5}{18}\e\right)\right]
 \label{gammagg}
\end{equation}
and provide the virtual contribution to the cross section
\begin{equation}
 \frac{\dif\sigma_{\pg\pb}}{\dif z_1\,\du\,\dd}=h_{\pg}^{(0)}(\ku)
 \delta(1-z_1)\dq\,2\Gamma_{\pg\pg}^{(+)}\,h_{\pb}^{(0)}(\kd)
 \quad.        \label{sggvir}
\end{equation}
The total fragmentation vertex for the gluon is then obtained by
summing Eqs.~(\ref{sggg}), (\ref{sqqg}) and (\ref{sggvir}), dropping
the common factor $h_{\pb}^{(0)}(\kd)$.

In order to extract the gluon impact factor, we have to eliminate the
$\log s$ term by subtracting the $1/z_1$ singularity coming from the
gluon splitting function (\ref{pgg}) by means of Eq.~(\ref{fauno}). Having
identified in Sec.~(\ref{ifnla}) $z_1$ as the momentum fraction of the
softer gluon in $\sf gb\rightarrow ggb$ scattering, to avoid
double counting due to final gluons identity, we restrict the naive
half phase space $z_1\in[\frac{q}{\sqrt{s}},1-\frac{k_1}{\sqrt{s}}]$
to $z_1\in[\frac{q}{\sqrt{s}},\frac{1}{2}]$. In so doing, the 1-loop
correction to the gluon impact factor, explicitly computed in
App.~\ref{integrali}, is given by
\begin{align}
 h_{\pg}^{(1)}(\kd)&=h_{\pg}^{(0)}(\kd)\,\o^{(1)}(\kd^2)\nonumber\\
 &\quad\left[\left(\frac{11}{6}-\frac{n_f}{3N_c}\right)+\left(\frac{11}{6}+
 \frac{(2+\e)n_f}{6N_c}\right)-\frac{C_F\,n_f}{N_c^2}\left(\frac{2}{3}+
 \frac{1}{3}\e\right)-\frac{\K}{N_c}\e\right]\label{hunog}
\end{align}
where $\K$ has been defined in Eq.~(\ref{k}).


\section{Collinear factorization and finite parts}\label{cffp}

We want now to analyze the structure of the quark and gluon impact
factors derived in Secs.~(\ref{qqcoll}) and (\ref{ggcoll}). The
explicit expressions reported in Eqs.~(\ref{hunoq}) and (\ref{hunog})
show the presence of
\begin{itemize}
\item a $\beta$-function coefficient $\ds{\frac{2\pi}{N_c}\,b=
      \frac{11}{6}-\frac{n_f}{3N_c}}$;
\item the finite part $\tilde{P}_{\pb\pa}(\o=0)$ in the $\o$ expansion
of the Mellin transform of the $\sf a\to b=q,g$ splitting functions
for the incoming parton {\sf a} under consideration, as depicted in
Tab.~(\ref{tabella});
\item a common and $\e$-finite factor $\ds{\frac{\K}{N_c}}$.
\end{itemize}
\newcommand{\slarga}{\raisebox{-6mm}{\rule{0mm}{14mm}}}
\begin{table}[b]
 \begin{tabular}{lll}%
 \hline%
 \vline~{\sf c$\,\leftarrow\,$a~}\vline&
 $ P_{\pc\pa}(z,\e)$\hfill\vline&
 $\int_0^1 z^{\o}P_{\pc\pa}(z,\e)\,\dif z$\hfill\vline\\
 \hline\hline%
\vline ~~\slarga{\sf q\hspace{3mm}q}\hfill\vline&%
\footnotesize $\ds{C_F\left[\left(\frac{1+z^2}{1-z}\right)_+
               -\e(1-z)+\frac{\e}{2}\,\delta(1-z)\right]}$\hfill\vline&%
\footnotesize $ 0+{\cal O}(\o)$\hfill\vline\\
 \hline%
\vline ~~\slarga{\sf g\hspace{3mm}q}\hfill\vline&%
\footnotesize $\ds{C_F\left[\frac{1+(1-z)^2}{z}+\e z\right]}$\hfill\vline&%
\footnotesize $\ds{\frac{2C_F}{\o}-C_F\left(\frac{3}{2}
         -\frac{1}{2}\e\right)+{\cal O}(\o)}$\hfill\vline\\
 \hline%
\vline ~~\slarga{\sf g\hspace{3mm}g}\hfill\vline&%
\footnotesize$\ds{2N_c\left[\frac{1-z}{z}+\left(\frac{z}{1-z}
        \right)_+ +z(1-z)-\frac{1}{12}\,\delta(1-z)\right]\;}$\hfill\vline&%
\footnotesize$\ds{\frac{2N_c}{\o}-\frac{11N_c+(2+\e)n_f}{6}
                    +{\cal O}(\o)}\quad$\hfill\vline\\
 \hline%
\vline ~~\slarga{\sf q\hspace{3mm}g}\hfill\vline&%
\footnotesize $\ds{n_f\left[1-\frac{2z(1-z)}{1+\e}\right]}$\hfill\vline&%
\footnotesize $\ds{n_f\left(\frac{2}{3}+\frac{1}{3}\,\e\right)
                     +{\cal O}(\o)}$\hfill\vline\\
 \hline%
\end{tabular}
\caption{\it Partonic splitting functions and Mellin transforms in
         $\o$ space.}
\label{tabella}
\end{table} 
This suggests writing the complete one-loop impact factors
$h_{\pa}(\kk)=h^{(0)}_{\pa}(\kk)+h^{(1)}_{\pa}(\kk)$ for both $\sf
a=q,g$ in the following manner:
\begin{align}
 h_{\pa}(\kk)&=h^{(0)}_{\pa}(\kk)+\frac{\o^{(1)}(\kk^2)}{N_c}\left[
 2\pi b\,h^{(0)}_{\pa}(\kk)-\sum_{\sf c}h_{\pc}^{(0)}(\kk)\tilde{P}_{\pc\pa}
 -h^{(0)}_{\pa}(\kk)\,\K\,\e\right]  \label{hak}\\
 &=\left[ 1-b\,\frac{\as}{\e}
 \left(\frac{\kk^2}{\mu^2}\right)^{\e}\,\right]\left[h^{(0)}_{\pa}(\kk)
 -\left(\sum_{\sf c}h_{\pc}^{(0)}(\kk)\tilde{P}_{\pc\pa}
 +h^{(0)}_{\pa}(\kk)\,\K\,\e\right)\right]+{\cal O}(\as^3)\;.\nonumber
\end{align}
The factor in front of the last expression provides the
renormalization of the coupling constant, and by introducing  the
one-loop running coupling
\begin{equation}
 \as(\kk^2)=\as\left[ 1-b\,\frac{\as}{\e}
 \left(\frac{\kk^2}{\mu^2}\right)^{\e}\right]\quad,
\end{equation}
it can be incorporated in $h_{\pa}^{(0)}$ itself.

Therefore, the impact factors in Eq.~(\ref{hak}) assume the form
\begin{equation}
 h_{\pa}(\kk^2)=h_{\pa}^{(0)}(\as(\kk^2))\left(1+\frac{\as}{2\pi}\,\K\right)
 +\frac{\as}{2\pi}\left[h_{\pq}^{(0)}\frac{\tilde{P}_{\pq\pa}}{\e}
 +h_{\pg}^{(0)}\frac{\tilde{P}_{\pg\pa}}{\e}\right]\left(\frac{\kk^2}{\mu^2}
 \right)^{\e}    \label{impatto}
\end{equation}
and thus satisfy the DGLAP equations with splitting functions
$\tilde{P}_{\pc\pa}(\o=0)$.

After factorization of the collinear singularities contained in the
latter, the only finite renormalization is the one implied by the
factor $\ds{\left(1+\frac{\as}{2\pi}\,\K\right)}$, which is universal,
i.e. independent of the parton type.

Several comments are in order. First, the collinear singularities in
$h_{\pa}$ do not contain the $\sim 1/\o$ terms of the splitting
functions, which have been subtracted out in the leading term. In this
context, we remark the difference between the double-$\kk$ cross
section defined here through $\kk$-factorization, and the
double-minijet inclusive cross section defined by Mueller and Navelet
\cite{mn87}.

In the latter case the cross section is inclusive over
all fragmentation products of the incoming partons which are not
identified, and measures the gluonic $\kk$'s only because of collinear
strong ordering.
Therefore, it factorizes the gluon structure function in the parton,
rather than the impact factor, with all its collinear singularities
included and, furthermore, can be applied only at leading $\log s$
level because of strong ordering.

On the other hand, in our case the definition of the $\kk$-dependent
cross section is more precise, but can be done theoretically rather
than experimentally, because one needs not only to identify all
fragmentation products, but also to subtract out the central region
tail~--~which appears to be hard to do experimentally. Because
of this difference, the collinear singularities to be factored out are
different in the two cases.

As a second point we remark that a universal renormalization, with the
same $\K$ coefficient, holds also for the soft part of the
one-loop timelike splitting functions \cite{fp80}.
This part has a next-to-leading
$1/(1-z_1)$ singularity, the leading one having a logarithmic
factor, and is therefore analogous to the impact factor, which
corresponds to the NL$x$ constant piece in the high-energy limit.

The above analogy is perhaps a hint \cite{kk96} to explain the universality
found here. However our result does not seem to follow in a clearcut
way, because of the difference between collinear and high-energy
factorization pointed out before.


\section{Discussion}\label{discus}

To sum up, we have defined here in a more explicit way the
$\kk$-factorization scheme introduced by one of us \cite{c98}, and we
have applied it to the calculation of the finite one-loop corrections
(Eq.~(\ref{impatto})) to the partonic impact factors, which turn out
to be universal. Furthermore, we have generalized the definition of
impact factors and of $\kk$-factorization scheme to hard processes
with colourless probes (Eqs.~(\ref{fadue}) and (\ref{hdue})) in such a
way as to be sure to avoid spurious infrared singularities in the
subtraction of the leading terms.

The above one-loop results, plus further use of $\kk$-factorization
and of the form (\ref{scomp}) of the gluon Green's function, allows to
extend the analysis of the NL$x$ $\kk$-factorization to two-loop level,
and to extract the NL$x$ BFKL kernel \cite{fl98,cc98}, for a factorized
scale of the type $s_0=k_1k_2$.

In order to close the NL$x$ calculations, the remaining task is to
justify Eq.~(\ref{fatt}) to all orders. In fact, checking
Eq.~(\ref{fatt}) at two-loops is not a real achievement, because it
only determines the NL$x$ kernel, depending on the scale choice.

The question is then whether the kernel iterates correctly at higher
orders for the class of factorized scales which is used. Investigating
this problem probably requires the use of group-theoretical variables
\cite{cdm69} in order to incorporate exact s-channel phase space in
the Mellin-transforms. It is hoped that this investigation will lead
to a better understanding of the role of $\kk$-dependent scales in the
NL$x$ kernel at higher orders as well.

\appendix

\section{Squared matrix elements in the fragmentation region}\label{elementi}

In this Appendix we report the explicit calculations of the $\sf gg^*$
fragmentation vertices, both for $\sf gg$ and for
$\sf q\overline{q}$ production. Let's remind that $p_j,\l_j$ and $a_j$
label respectively the momentum, the polarization and the colour
quantum number of the $j$-th external particle.

\subsection{Polarization vectors}

Here we present the conventions for the polarization vectors of the
gluons that we have adopted in this paper. Even if it would be simpler
to use real polarization vectors, we consider complex ones that in the
$D\to 4$ limit reduce to the usual helicity basis.

The light-cone coordinates are defined to be
\begin{equation*}
 x^{(\pm)}=\frac{x^0\pm x^3}{\sqrt{2}}\quad;\quad
 x^{\mu}=(x^0,x^1,x^2,x^3;x^{\e})=\langle x^+,x^-;x^1,x^2;x^{\e}\rangle
\end{equation*}
where $x^{\e}$ represents the extra coordinates in a $2\e$ dimensional
space. We introduce also the complex transverse coordinate
$\tilde{x}=x^1+ix^2$.

There are $2+2\e$ physical polarization vectors and we take, in the
$p_2$ gauge,
\begin{equation*}
 \po_{(+)}(q)=\poc_{(-)}(q)=\frac{1}{\sqrt{2}}\langle0,\frac{\tilde{q}}{q^+};
 1,i;0\rangle\quad;\quad\po_{(\eta)}(q)=\langle0,0;0,0;\hat{\eta}\rangle
\end{equation*}
where $\hat{\eta}$ is the $\eta$-th unit-vector in the
$2\e$-dimensional extra space: $\eta\in\{1,\cdots,2\e\}$.
Indicating with $\l$ any of the physical polarization,
$\l\in\{+,-,1,\cdots,2\e\}$ we have
\begin{equation*}
 \po_{(\l)}(q)\ps q=0\quad;\quad\po_{(\l)}(q)\ps p_2=0\quad
 \forall\l=+,-,\eta\quad.
\end{equation*}
Defining the involution $\l\mapsto\l^*$ acting on the polarization
indices in the following way:
\begin{equation*}
 (+)^*=(-)\quad;\quad(-)^*=(+)\quad;\quad(\eta)^*=(\eta)\quad\Rightarrow
 \quad\poc_{(\l)}=\po_{(\l^*)}\quad,
\end{equation*}
the orthonormality relations assume the form:
\begin{align*}
 &\poc_{(\l_a)}(q_a)\ps\po_{(\l_b)}(q_b)=-\delta_{\l_a\l_b}\\
 &\po_{(\l_a)}(q_a)\ps\po_{(\l_b)}(q_b)=
 \poc_{(\l_a)}(q_a)\ps\poc_{(\l_b)}(q_b)=-\delta_{\l_a\l_b^*}=
 -\delta_{\l_a^*\l_b}\quad\forall q_a,q_b\quad.
\end{align*}
The polarization vectors $\{\po_{(\l)}(q):\l=+,-,\eta\}$ are a basis
for the subspace orthogonal to the plane $\langle p_2,q\rangle$ and
\begin{equation}
 -\sum_{\l}\poc\,_{(\l)}^{\alpha}(q)\po_{(\l)}^{\beta}(k)=g_{\alpha\beta}
 -\frac{n_{\alpha}q_{\beta}}{n\ps q}-\frac{k_{\alpha}n_{\beta}}{n\ps k}
 +\frac{q\ps k}{n\ps q\,n\ps k}n_{\alpha}n_{\beta}
 \equiv\Delta_{\alpha\beta}^{(n)}(q,k)  \label{proiet}
\end{equation}
where n is the gauge vector ($n=p_2$ for us). Contracting the 
Lorentz indices we obtain
\begin{equation*}
 \Delta_{\alpha}^{\;\alpha}(q,k)=-(2+2\e)\quad.
\end{equation*}

\subsection{$\sf gb\to ggb$ squared matrix element}

The transition amplitude for gluon-parton scattering with production
of an additional gluon in the fragmentation region of the incoming
gluon {\sf a} (Fig.~(\ref{diagggg})) in the high energy limit
$s\to\infty$ has been given in Eq.~(\ref{aggg}) in terms of the
fragmentation tensor (\ref{amumu})
$A_{\mu_1\mu_3\mu_5}^{a_1a_3a_5b}\equiv
 A_{\mu_1\mu_3\mu_5\nu}^{a_1a_3a_5b}\,2p_2^{\nu}$ and the current
(\ref{d}). By using physical polarization $\po\ps p_2=0$ we can drop
the $p_2^{\mu}$-component of $D^{\mu}$ and retain
\begin{equation*}
  D^{\mu}(x,y,z)=\frac{1}{x\ps y}\left[\frac{p_2\ps y}{x\ps z}
 (y\ps z-p_2\ps p_4)x^{\mu}+(x\ps p_2)y^{\mu}\right]\quad.
\end{equation*}
Applying in Eq.~(\ref{amumu}) the Jacobi identity
\begin{equation*}
 \f^{a_3a_5e}\f^{ea_1b}=\f^{a_1a_5e}\f^{ea_3b}-
 \f^{a_1a_3e}\f^{ea_5b}\quad,
\end{equation*}
the fragmentation tensor reads
\begin{align*}
 A_{\mu_1\mu_3\mu_5}^{a_1a_3a_5b}=&4g^2\Big\{\f^{a_1a_3e}\f^{ea_5b}
 \Big[g_{\mu_1\mu_5}
 \big(D_{\mu_3}(-p_1,p_5,p_3)+D_{\mu_3}(p_5,-p_1,p_3)\big)\\
 &\hspace{20mm}-g_{\mu_1\mu_3}D_{\mu_5}(p_3,-p_1,p_5)-
 g_{\mu_3\mu_5}D_{\mu_1}(p_3,p_5,-p_1)\Big]
 \;+\;\big(3\leftrightarrow5)\Big\}\\
 =&4g^2\Big\{\f^{a_1a_3e}\f^{ea_5b}\Big[g_{\mu_1\mu_5}D^{A}_{\mu_3}
 +g_{\mu_1\mu_3}D^{B}_{\mu_5}+g_{\mu_3\mu_5}D^{C}_{\mu_1}
 \;+\;\big(3\leftrightarrow5\big)\Big\}
\end{align*}
where, for brevity, we have defined
\begin{align*}
 D_{A}^{\mu}&=D^{\mu}(-p_1,p_5,p_3)+D^{\mu}(p_5,-p_1,p_3)\quad,\\
 D_{B}^{\mu}&=D^{\mu}(p_3,-p_1,p_5)\quad,\\
 D_{C}^{\mu}&=D^{\mu}(p_3,p_5,-p_1)\quad.
\end{align*}
By averaging over all colours and polarizations, we obtain the
unpolarized emission probability
\begin{equation*}
 P_{\pg\pg\pb}=g^2\,\frac{C_{\pb}}{(N_c^2-1)^2}\,
 \frac{1}{2+2\e}\,\frac{1}{(k_2^2)^2}
 \sum_{a_j,b,\l_j}\left|\po_1^{\mu_1}\,\poc\,_3^{\mu_3}\,\poc\,_5^{\mu_5}
 \;A_{\mu_1\mu_3\mu_5}^{a_1a_3a_5\,b}\right|^2\quad.
\end{equation*}
Now we isolate colour and polarization structures by defining
\begin{align*}
 Z_{135}&=Z_{\l_1\l_3\l_5}(p_1,p_3,p_5)\equiv
 \po_1^{\mu_1}\,\poc\,_3^{\mu_3}\,\poc\,_5^{\mu_5}\left(
 g_{\mu_1\mu_5}D^{A}_{\mu_3}+g_{\mu_1\mu_3}D^{B}_{\mu_5}+
 g_{\mu_3\mu_5}D^{C}_{\mu_1}\right)\\
 &=-\delta_{\l_1\l_5}\poc_3\ps D_{A}+\delta_{\l_1\l_3}\poc_5\ps D_{B}
 +\delta_{\l_3\l_5^*}\,\poc_3\ps D_{C}
\end{align*}
in such a way that
\begin{equation*}
 \po_1^{\mu_1}\,\poc\,_3^{\mu_3}\,\poc\,_5^{\mu_5}
 \;A_{\mu_1\mu_3\mu_5}^{a_1a_3a_5\,b}=
 4g^2\left[\f^{a_1a_3e}\f^{ea_5b}Z_{135}
 \;+\;\big(3\leftrightarrow5\big)\right]
\end{equation*}
and
\begin{align}
 P&=\frac{16g^6}{(2+2\e)}\,\frac{C_{\pb}}{(N^2_c-1)^2}\,\frac{1}{(k_2^2)^2}
 \sum_{a_j,b,\l_j}\left|\f^{a_1a_3e}\f^{ea_5b}Z_{135}+
 \f^{a_1a_5e}\f^{ea_3b}Z_{153}\right|^2\nonumber\\
 &=\frac{16g^6}{2+2\e}\,\frac{C_{\pb}N_c^2}{N^2_c-1}\,\frac{1}{(k_2^2)^2}
 \sum_{\l_j}\left[|Z_{135}|^2+|Z_{153}|^2+\frac{1}{2}(Z_{135}^*Z_{153}+
 Z_{153}^*Z_{135})\right]\quad.      \label{ugo}
\end{align}
We evaluate the terms in square brackets by using Eq.~(\ref{proiet}).
In this way we get
\begin{align}
 \hspace{-3mm}\sum_{\l_j}|Z_{135}|^2=&-(2+2\e)\!\left[
 \Delta_{\alpha\beta}(p_3,p_3)D_{A}^{\alpha}D_{A}^{\beta}+
 \Delta_{\alpha\beta}(p_5,p_5)D_{B}^{\alpha}D_{B}^{\beta}+
 \Delta_{\alpha\beta}(p_1,p_1)D_{C}^{\alpha}D_{C}^{\beta}\right]\nonumber\\
 &+2\left[\Delta_{\alpha\beta}(p_3,p_5)D_{A}^{\alpha}D_{B}^{\beta}+
 \Delta_{\alpha\beta}(p_3,p_1)D_{A}^{\alpha}D_{C}^{\beta}-
 \Delta_{\alpha\beta}(p_1,p_5)D_{C}^{\alpha}D_{B}^{\beta}
 \right]\quad.     \label{espa}
\end{align}
By expanding the Lorentz contractions substituting the appropriate
kinematical variables, the interference terms (i.e. the second line of
Eq.~(\ref{espa})) cancel each other, and the first line yields
\begin{equation}
 \sum_{\l_j}|Z_{135}|^2=\frac{(1+\e)s^2\kd^2(1-z_1)^2[1+z_1^4+(1-z_1)^4]
 }{\ku^2\,(z_1\ku+(1-z_1)\qq)^2}\quad. \label{utc}
\end{equation}
The second term in (\ref{ugo}) has the same structure as the first
one, and gives
\begin{align}
 \sum_{\l_j}|Z_{153}|^2&=\sum_{\l_j}\left.|Z_{135}|^2
 \right|_{p_3\leftrightarrow p_5}=\sum_{\l_j}\left.|Z_{135}|^2
 \right|_{\substack{z_1\leftrightarrow1-z_1\\ 
                    \qq\leftrightarrow-\ku}}\nonumber\\
 &=\frac{(1+\e)s^2\kd^2\,z_1^2[1+z_1^4+(1-z^1)^4]}{%
 \qq^2\,(z_1\ku+(1-z_1)\qq)^2} \quad.\label{uct}
\end{align}
For the third term in (\ref{ugo}) we obtain
\begin{align*}
 \hspace{-3mm}\sum_{\l_j}Z_{135}^*Z_{153}=&(2+2\e)\!\left[
 \Delta_{\alpha\beta}(p_3,p_3)D_{A}^{\alpha}\overline{D}_{B}^{\beta}+
 \Delta_{\alpha\beta}(p_5,p_5)D_{B}^{\alpha}\overline{D}_{A}^{\beta}-
 \Delta_{\alpha\beta}(p_1,p_1)D_{C}^{\alpha}\overline{D}_{C}^{\beta}
 \right]\nonumber\\
 &-\Delta_{\alpha\beta}(p_3,p_5)D_{A}^{\alpha}\overline{D}_{A}^{\beta}+
 \Delta_{\alpha\beta}(p_3,p_1)D_{A}^{\alpha}\overline{D}_{C}^{\beta}-
 \Delta_{\alpha\beta}(p_5,p_3)D_{B}^{\alpha}\overline{D}_{B}^{\beta}\\
 &- \Delta_{\alpha\beta}(p_5,p_1)D_{B}^{\alpha}\overline{D}_{C}^{\beta}+
 \Delta_{\alpha\beta}(p_1,p_5)D_{C}^{\alpha}\overline{D}_{A}^{\beta}-
 \Delta_{\alpha\beta}(p_1,p_3)D_{C}^{\alpha}\overline{D}_{B}^{\beta}\quad.
\end{align*}
Also in this case the interference terms cancel each other, and we
obtain
\begin{equation}
 \sum_{\l_j}Z_{135}^*Z_{153}=-\frac{(1+\e)s^2\kd^2\,\ku\ps\qq\,z_1(1-z_1)
 [1+z_1^4+(1-z_1)^4]}{\ku^2\,\qq^2\,(z_1\ku+(1-z_1)\qq)^2}\quad.
 \label{utcuct}
\end{equation}
The fourth and last term in (\ref{ugo}) is the complex conjugate of
the third one and hence, since (\ref{utcuct}) is real, they are equal.

Multiplying Eq.~(\ref{ugo}) by the three-body phase space
(\ref{sftre}) and substituting the expressions
(\ref{utc},\ref{uct},\ref{utcuct}), we finally obtain the differential
cross section (\ref{sggg}) for $\sf gb\to ggb$ scattering.

\subsection{$\sf gb\to q\overline{q}b$ squared matrix element}

The calculation of the $\sf gb\to q\overline{q}b$ cross section is
analogous to that of $\sf qb\to qgb$ and requires the evaluation of
the $\sf gg^*\to q\overline{q}$ vertex, which can be obtained by means of
$\kk$-factorization \cite{cch90}:
\begin{equation*}
 \overline{|\M_{ gg^*\to q\overline{q}}|^2}=
 \frac{1}{(2+2\e)(N_c^2-1)}\sum_{a,b,\l_j}\left|\po^{\mu}\,
 \overline{u}_3 B_{\mu\nu}^{ab}v_5\,2p_2^{\nu}\right|^2\quad.
\end{equation*}
Adopting the Feynman gauge,
\begin{align*}
 B_{\mu\nu}^{ab}=g^2&\Big\{[T^a,T^b]\frac{1}{2p_3p_5}\big((p_1+2k_2)_{\mu}
 \gamma_{\nu}+(\ta{p}_1-\ta{k}_2)g_{\mu\nu}-(2p_1+k_2)_{\nu}\gamma_{\mu}
 \big)\\
 &\;+T^aT^b\gamma_{\nu}\,\frac{(\ta{p}_1-\ta{q})}{2p_1q}\,\gamma_{\mu}
 +T^bT^a\gamma_{\mu}\,\frac{(\ta{p}_3-\ta{p}_1)}{2p_1p_3}\,\gamma_{\nu}
 \Big\}
\end{align*}
and by using physical polarization with $\po\ps p_2=0$ and the high
energy kinematics, we can write the colour decomposition
\begin{equation*}
 B_{\mu\nu}^{ab}\po^{\mu}2p_2^{\nu}\equiv B^{ab}=
 \{T^a,T^b\}B^{(+)}+[T^a,T^b] B^{(-)}\quad,
\end{equation*}
where
\begin{align*}
 B^{(\pm)}&=b_1^{(\pm)}\left(\ta{p}_2\po\ps q-\frac{1}{2}\ta{\po}
    \ta{p}_2\ta{p}_1\right)+b_2^{(\pm)}\left(\ta{p}_2\po\ps p_3-
    \frac{1}{2}\ta{\po}\ta{p}_1\ta{p}_2\right)\\
 &b_1^{(+)}=-\frac{1}{p_1q}&b_2^{+}&=\frac{1}{p_1p_3}\\
 &b_1^{(-)}=-\frac{1}{p_1q}+\frac{2}{p_3q}&
 b_2^{-}&=-\frac{1}{p_1p_3}+\frac{2}{p_3q}\quad.
\end{align*}
By performing the gamma matrices algebra and the polarization sum in
$D=4+2\e$ dimensions, we obtain, for each colour structure,
\begin{align*}
 \sum_{\l_j}|\overline{u}_3Bv_5|^2=&(2+2\e)s^2\P_{\pq\pg}(z_1,\e)\times\\
 &\times\left\{b_1^2(1-z_1)p_1q+b_2^2z_1p_1p_3+
               b_1b_2[p_3q-z_1p_1p_3-(1-z_1)p_1q]\right\}
\end{align*}
where $\P_{\pq\pg}$ is given in Eq.~(\ref{pqg}) and is related to the
gluon to quark splitting function. By averaging the square of the
amplitude (\ref{aqqg}) over all colours and polarizations, the
emission probability reads
\begin{equation*}
 P=\overline{|\M_{ gg^*\to q\overline{q}}|^2}\,\frac{1}{(k_2^2)^2}\,
 \frac{C_{\pb}}{N_c^2-1}
\end{equation*}
which, multiplied by the three-body phase space (\ref{sftre}), gives
the differential cross section (\ref{sqqg}) for
$\sf gb\to q\overline{q}b$ scattering.


\section{Calculation of $\ku$ integrals}\label{integrali}

In this Appendix we show the relevant integrations (over $z_1$ and in $\ku$
transverse space) of the fragmentation
vertex in order to get the 1-loop impact factors, both for the quark
and for the gluon case.

\subsection{Quark impact factor}
 
In first place, we show that the Sudakov term  of {\sf qgb}
production, upon $\ku$ and $z_1$ integration, cancels out with the
$C_F$ term of the virtual correction. In fact, the $C_F$ contribution
to the fragmentation vertex $F_{\pq}(z_1,\ku,\kd)$ is
\begin{align*}
 h_{\pq}^{(0)}(\ku)&\,\frac{\P_{\pg\pq}(z_1,\e)}{\Gamma(1-\e)\mu^{2\e}}\,
 \frac{C_F\as}{\pi}\,\frac{z_1^2\,\ku^2}{\qq^2(\qq-z_1\kd)^2}\\
 &=\sqrt{\frac{\pi}{N_c^2-1}}\frac{2C_F\as N_{\e}}{\mu^{2\e}}
 \frac{C_F\as}{2\pi\Gamma(1-\e)\mu^{2\e}}\,z_1\left[1+(1-z_1)^2+
 \e z_1^2\right]\frac{1}{\qq^2(\qq-z_1\kd)^2}.
\end{align*}
In order to perform the $\ku$-integration, we use the formula
\begin{equation}
 \int\frac{\dk}{(\kk ^{2})^{1-\alpha}(({\boldsymbol p}-\kk)^2)^{1-\beta}}=
 \frac{\Gamma(1-\alpha-\beta-\e)\Gamma(\alpha+\e)
 \Gamma(\beta+\e)}{\Gamma(\alpha+\beta+2\e)\Gamma(1-\alpha)\Gamma(1-\beta)}
 \,({\boldsymbol p}^2)^{\alpha+\beta+\e-1}   \label{formula}
\end{equation}
to obtain
\begin{equation*}
 \int\frac{\du}{\qq^2(\qq-z_1\kd)^2}
 =\int\frac{\dif[\qq]}{\qq^2(\qq-z_1\kd)^2}
 =\frac{\Gamma(1-\e)\Gamma^2(\e)}{\Gamma(2\e)}\,(\kd^2)^{\e-1}\,z_1^{2\e-2}
 \quad.
\end{equation*}
The integral over $z_1$, due to the vanishing of the integrand when
$z_1$ goes to $0$, can be extended in its lower limit from
$q/\sqrt{s}$ -- the half phase space -- down to $0$,
introducing in this way a negligible error in the limit $s\to\infty$.
This yields
\begin{align*}
 \sqrt{\frac{\pi}{N_c^2-1}}\frac{2C_F\as N_{\e}}{\mu^{2\e}}\,
 &\frac{C_F\as}{2\pi\mu^{2\e}}\,(\kd^2)^{\e-1}
 \frac{\Gamma(1-\e)\Gamma^2(\e)}{\Gamma(2\e)}\int_0^1\dif z_1\,z_1^{2\e-1}
 \left[1+(1-z_1)^2+\e z_1^2\right]\\
 &=h_{\pq}^{(0)}(\kd)\,\frac{C_F\as}{\pi}\,
 \frac{\Gamma^2(1+\e)}{\e\,\Gamma(1+2\e)}\left(\frac{\kd^2}{\mu^{2\e}}
 \right)^{\e}\left(\frac{1}{\e}-
 \frac{2}{1+2\e}+\frac{1}{2}\right)\\
 &=h_{\pq}^{(0)}(\kd)\,2\o^{(1)}(\kd^2)\frac{C_F}{N_c}\left(
 -\frac{1}{\e}+\frac{3}{2}-4\e\right)+{\cal O}(\e)
\end{align*}
which exactly cancels the part of the virtual term proportional to
$C_F$
\begin{equation*}
 \int\du\int_0^1\dif z_1\,\left.h_{\pq}^{(0)}(\ku)\,2\Gamma_{\pq\pq}^{(+)}
 \right|_{C_F}\delta(1-z_1)\dq=h_{\pq}^{(0)}(\kd)\,2\o^{(1)}(\kd^2)
 \left.\Gamma_{\pq\pq}^{(+)}\right|_{C_F}
\end{equation*}
as can be seen with a glance to Eq.~(\ref{gammaqq}).

The effective quark fragmentation vertex is then as given in
Eq.~(\ref{effeq}), i.e.
\begin{align*}
 \hspace{-9mm}F_{\pq}(z_1,\ku,\kd)=h^{(0)}_{\pq}(\ku)&
  \left\{\dq\delta(1-z_1)2\o^{(1)}(\ku^2)\left[
  \frac{11}{12}-\left(\frac{85}{36}+\frac{\pi^2}{4}\right)\!\e-
  \frac{n_f}{N_c}\left(\frac{1}{6}-\frac{5}{18}\e
  \right)\right]\right.   \\
  &\,\left.+\frac{\P_{\pg\pq}(z_1,\e)}{\Gamma(1-\e)\mu^{2\e}}\,
  \frac{\ab}{\qq^2}
  \,\frac{(1-z_1)\qq\ps(\qq-z_1\kd)}{(\qq-z_1\kd)^2}\right\}\quad.
\end{align*}
In order to obtain the 1-loop correction to the quark impact factor,
we have to use Eq.~(\ref{fauno}) to subtract the $1/z_1$
singular part, which gives the $\log s$ behaviour of the
cross section. In this way we get
\begin{align}
 \hspace{-2mm}\ab F_{\pq}^{(1)}&(z_1,\ku,\kd)
 =\sqrt{\frac{\pi}{N_c^2-1}}\,2C_F\as N_{\e}\times   \label{funoq}\\
 &\left\{\dq\delta(1-z_1)\frac{2\o^{(1)}(\ku^2)}{\ku^2}
 \left[\frac{11}{12}-\left(\frac{85}{36}+\frac{\pi^2}{4}\right)\e-
 \frac{n_f}{N_c}\left(\frac{1}{6}-\frac{5}{18}\e\right)\right]
 +\right.\nonumber\\
 +&\left.\frac{\ab}{\Gamma(1-\e)\mu^{2\e}}\left[\frac{1}{2z_1}
 \big(1+(1-z_1)^2+\e z_1^2\big)
 \frac{(1-z_1)\qq\ps(\qq-z_1\kd)}{\ku^2\,\qq^2\,(\qq-z_1\kd^2)}-
 \frac{\Theta(q-z_1k_1)}{z_1\,\ku^2\,\qq^2}\right]\right\}\nonumber
\end{align}
and then we have to integrate over $z_1\in[0,1]$ and $\ku$ at fixed
$\kd$.

The integration of the first term (i.e. the virtual contribution) is
straightforward and yields
\begin{equation}
 h_{\pq}^{(0)}(\kd)\,\o^{(1)}(\kd^2)\left[\frac{11}{6}-
 \left(\frac{85}{18}+\frac{\pi^2}{2}\right)\e-\frac{n_f}{N_c}
 \left(\frac{1}{3}-\frac{5}{9}\e\right)\right]\quad. \label{hqa}
\end{equation}
Using the identity $\Theta(z_1)=1-\Theta(-z_1)$ the last line of
Eq.~(\ref{funoq}) yields the multiple integral
\begin{align}
 \hspace{-3mm}I_{\pq}=\frac{\ab}{\Gamma(1-\e)}&\left\{\int_0^1\dif z_1
 \left[\P_{\pg\pq}(z_1,\e)\,
 (1-z_1)\int\du\,\frac{\qq\ps(\qq-z_1\kd)}{\ku^2\,\qq^2\,(\qq-z_1\kd)^2}-
 \frac{1}{z_1}\int\frac{\du}{\ku^2\,\qq^2}\right]\right.\nonumber \\
 &\,\left.+\int\du\log\frac{k_1}{q}
 \Theta_{k_1q}\,\frac{1}{\ku^2\,\qq^2}\right\}\quad.   \label{iq}
\end{align}
It is convenient to perform in first place the $\ku$ integration, since
dimensional regularization provides simple integrals in the transverse
momentum space

The integrand of the first $\ku$-integral in (\ref{iq}) can be written
as a sum of terms with two factors in the denominator:
\begin{align*}
 \hspace{-8.3mm}\frac{\qq\ps(\qq-z_1\kd)}{\ku^2\,\qq^2\,(\qq-z_1\kd)^2}&=
 \frac{1}{\ku^2\,\qq^2\,(\qq-z_1\kd)^2}\left[\frac{1-z_1}{2}\qq^2-
 \frac{z_1^2}{2(1-z_1)}\ku^2+\frac{1}{2(1-z_1)}(\qq-z_1\kd)^2\right] \\
 &=\frac{1}{2(1-z_1)}\left[\frac{(1-z_1)^2}{\ku^2\,(\qq-z_1\kd)^2}-
 \frac{z_1^2}{\qq^2\,(\qq-z_1\kd)^2}+\frac{1}{\ku^2\,\qq^2}\right]\quad,
\end{align*}
thus allowing, with an appropriate shift in the integration variable,
to apply (\ref{formula}) with $\alpha=\beta=0$ and to obtain
\begin{equation*}
 \int\du\frac{\qq\ps(\qq-z_1\kd)}{\ku^2\,\qq^2\,(\qq-z_1\kd)^2}=
 \frac{\Gamma(1-\e)\Gamma^2(1+\e)}{\e\;\Gamma(1+2\e)}\,(\kd^2)^{\e-1}
 \frac{1-z_1^{2\e}+(1-z_1)^{2\e}}{1-z_1}\quad.
\end{equation*}

The second $\ku$-integral is immediately solved:
\begin{equation*}
 \int\frac{\du}{\ku^q\,\qq^2}=
 \frac{\Gamma(1-\e)\Gamma^2(1+\e)}{e\;\Gamma(1+2\e)}\,2(\kd^2)^{\e-1}
 \quad.
\end{equation*}

For the third integral we use the representation
\begin{equation*}
 \log\frac{a}{b}\,\Theta_{ab}=\lim_{\alpha\to0^+}
 \int_{-i\infty}^{+i\infty}\frac{\dif\l}{2\pi i}\,\frac{1}{(\l+\alpha)^2}
 \left(\frac{a}{b}\right)^{\l}\quad,
\end{equation*}
valid for $a,b>0$, which allows us to write
\begin{align*}
 \int\du\,&\log\frac{k_1}{q}\,\Theta_{k_1q}\,\frac{1}{\ku^2\,\qq^2}=
 \frac{1}{2}\lim_{\alpha\to 0^+}\int_{-i\infty}^{+i\infty}
 \frac{\dif\l}{2\pi i}\,\frac{1}{(\l+\alpha)^2}\int\du\left(
 \frac{\ku^2}{\qq^2}\right)^{\l}\,\frac{1}{\ku^2\,\qq^2}\\
 &=\frac{1}{2}\lim_{\alpha\to 0^+}\int_{-i\infty}^{+i\infty}
 \frac{\dif\l}{2\pi i}\,\frac{1}{(\l+\alpha)^2}\frac{\Gamma(1-\e)
 \Gamma(\l+\e)\Gamma(-\l+\e)}{\Gamma(2\e)\Gamma(1-\l)\Gamma(1+\l)}
 \,(\kd^2)^{\e-1}\quad.
\end{align*}
The function
\begin{equation}
 \l\mapsto\frac{\Gamma(\l+\e)\Gamma(-\l+\e)}{\Gamma(2\e)\Gamma(1-\l)
 \Gamma(1+\l)\,(\l+\alpha)^2}   \label{funz}
\end{equation}
rapidly vanishes for $|\l|\rightarrow\infty$ in all directions, apart for
the two real semi-axes. Therefore, we can displace the integration
contour aroun the positive real semi-axis, enclosing the poles placed
in $\l=\e+n\;:n\in\mathbb{N}$. The residue of (\ref{funz}) in
$\l=\e+n$ is
\begin{equation*}
 \frac{(-1)^{n+1}}{n!}\,\frac{\Gamma(n+2\e)}{\Gamma(1+n+\e)\Gamma(1-n-\e)
 (n+\alpha+\e)^2}\quad.
\end{equation*}
In this way we can set $\alpha=0$ and replace the integral with the sum
\begin{equation*}
 \frac{1}{2}\frac{\Gamma(1-\e)}{\Gamma(2\e)}(\kd^2)^{\e-1}\sum_{n=0}^{\infty}
 \frac{(-1)^n}{n!}\,\frac{\Gamma(n+2\e)}{\Gamma(1+n+\e)\Gamma(1-n-\e)
 (n+\e)^2}\quad.
\end{equation*}
The first term in the sum ($n=0$) is
\begin{equation*}
 \frac{\Gamma(2\e)}{\Gamma(1+\e)\Gamma(1-\e)\,\e^2}\quad,
\end{equation*}
which is ${\cal O}(\e^{-3})$ and contribute to the integral to
${\cal O}(\e^{-2})$. The other terms ($n \geqslant 1$) are
${\cal O}(\e^2)$ and contribute to the integral to ${\cal O}(\e)$.
In conclusion
\begin{equation}
 \int\du\,\log\frac{k_1}{q}\,\Theta_{k_1q}\,\frac{1}{\ku^2\,\qq^2}=
 \frac{\Gamma(1-\e)\Gamma^2(1+\e)}{\e\;\Gamma(1+2\e)}\,
 \frac{1}{2\e}\,(\kd^2)^{\e-1}+{\cal O}(\e)\quad.
\end{equation}

We are left with the $z_1$ integration:
\begin{align*}
 I_{\pq}&=\ab\,\frac{\Gamma^2(1+\e)}{\e\,\Gamma(1+2\e)}\,(\kd^2)^{\e-1}\times\\
 &\quad\times\left\{\int_0^1\dif z_1\left[\frac{1}{2z_1}
 \bigg(1+(1-z_1)^2+\e z_1^2\bigg)
 \bigg(1-z_1^{2\e}+(1-z_1)^{2\e}\bigg)-\frac{2}{z_1}
 \right]+\frac{1}{2\e}\right\}\quad.
\end{align*}
We observe that the $1/z_1$ singularity  coming from
$\P_{\pg\pg}$ is correctly subtacted out, and the $1/2\e$ term
produced by the $\log(k_1/q)$ integration cancels out with
\begin{align*}
 &\int_0^1\dif z_1\left[\frac{1}{2z_1}\bigg(1+(1-z_1)^2+\e z_1^2\bigg)
 \bigg(1-z_1^{2\e}+(1-z_1)^{2\e}\bigg)-\frac{2}{z_1}\right]=\\
 &\quad-\frac{1}{2\e}+\psi(1)-\psi(1+2\e)-1+\frac{1+\e}{2}\left(\frac{1}{2}+
 \frac{1}{1+2\e}-\frac{1}{1+\e}\right)\quad.
\end{align*}
Summing the virtual contribution (\ref{hqa}) and
$2C_F\as N_{\e}\mu^{-2\e}I_{\pq}$ yields finally
\begin{align*}
 \ab h_{\pq}^{(1)}(\kd)
 &=\int\du\int_0^1\dif z_1\,\ab F_{\pq}^{(1)}(z_1,\ku,\kd)\\
 &=h_{\pq}^{(0)}(\kd)\o^{(1)}(\kd^2)\left[\left(\frac{11}{6}-
 \frac{n_f}{3N_c}\right)+\left(\frac{3}{2}-\frac{1}{2}\e\right)-
 \left(\frac{67}{18}-\frac{\pi^2}{6}-\frac{5n_f}{9N_c}\right)\e\right]
\end{align*}
as given in Eq.~(\ref{hunoq}) of the text.

\subsection{Gluon impact factor}

The subtracted fragmentation vertex $F_{\pg}^{(1)}$ for the gluon,
calculated in Sec.~(\ref{ggcoll}), is
\begin{align*}
 \ab F_{\pg}^{(1)}&(z_1,\ku,\kd)=\sqrt{\frac{\pi}{N_c^2-1}}
 \frac{2C_A\as N_{\e}}{\mu^{2\e}}\times\\
 \times&\left\{\dq\delta(1-z_1)\frac{2\o^{(1)}(\ku^2)}{\ku^2}\left[
 \frac{1}{\e}-\frac{11}{12}+\left(\frac{67}{36}-\frac{\pi^2}{4}\right)\e
 +\frac{n_f}{N_c}\left(\frac{1}{6}-\frac{5}{18}\e\right)\right]\right.\\
 &+n_f\,\frac{\P_{\pq\pg}(z_1,\e)}{4\Gamma(1-\e)\mu^{2\e}}
 \left[\frac{\as}{\pi}\,
 \frac{-z_1(1-z_1)\ku\ps\qq}{\ku^2\,\qq^2\,(z_1\ku+(1-z_1)\qq)^2}
 +\frac{C_F\as}{C_A\pi}\,\frac{1}{\ku^2\,\qq^2}\right]\\
 &+\frac{\ab}{\Gamma(1-\e)\mu^{2\e}}\,\Theta(\frac{1}{2}-z_1)\P_{\pg\pg}(z_1)\,
 \frac{z_1^2\ku^2+(1-z_1)^2\qq^2+z_1(1-z_1)\ku\ps\qq}{%
 \ku^2\;\qq^2\;(z_1\ku+(1-z_1)\qq)^2}\\
 &-\left.\frac{\ab}{\Gamma(1-\e)\mu^{2\e}}\,
 \frac{\Theta(q-z_1 k_1)}{z_1\,\ku^2\,\qq^2}\right\}\quad.
\end{align*}
Let's now integrate over $z_1$ and $\ku$ in order to obtain the one-loop
correction to the gluon impact factor.

The first line in curly brackets, corresponding to the virtual
contribution, yields immediately
\begin{equation}
 h_{\pg}^{(0)}(\kd)\,\o^{(1)}(\kd^2)\left[
 \frac{2}{\e}-\frac{11}{6}+\left(\frac{67}{18}-\frac{\pi^2}{2}\right)\e
 +\frac{n_f}{N_c}\left(\frac{1}{3}-\frac{5}{9}\e\right)\right]\quad.
 \label{hga}
\end{equation}

The second line corresponds to the {\sf q$\overline{\sf q}$b} final
state. Since it is finite in the limit $z_1\rightarrow0$, we may extend
the $z_1$ integration from the half phase space
$z_1\in[\frac{q}{\sqrt{s}},1]$ down to $z_1=0$ introducing in this way a
negligible error in the $s\rightarrow\infty$ limit. To perform the $\ku$
integration, we decompose the bigger fraction in a sum containing
terms which have only two factors in the denominator, allowing us
to apply Eq.~(\ref{formula}) and to obtain
\begin{equation}
 h_{\pg}^{(0)}(\kd)\,\o^{(1)}(\kd^2)\left[\frac{n_f}{N_c}\left(
 -\frac{1}{3}+\frac{23}{18}\e\right)-\frac{n_f\,C_F}{N_c^2}\left(
 \frac{2}{3}+\frac{1}{3}\e\right)\right]\quad. \label{hgb}
\end{equation}

The third and the fourth line represent the {\sf ggb}
contribution and the central region subtraction respectively. Let's
define, for brevity, the function
\begin{equation*}
 R(z_1,\ku,\qq)\equiv\frac{z_1^2\ku^2+(1-z_1)^2\qq^2+z_1(1-z_1)\ku\ps\qq}{%
 \ku^2\;\qq^2\;(z_1\ku+(1-z_1)\qq)^2}\quad.
\end{equation*}
Using the symmetry properties
\begin{equation*}
 R(1-z_1,\ku,\qq)=R(z_1,\qq,\ku)=R(z_1,-\qq,-\ku)
\end{equation*}
and Eq.~(\ref{pgg}) which allows us to write
\begin{equation*}
 \P_{\pg\pg}(z_1)=\pp(z_1)+\pp(1-z_1)\quad;\quad\pp(z_1)=\left(\frac{1}{z_1}+
 \frac{z_1}{2}\right)(1-z_1)\quad,
\end{equation*}
we have:
\begin{align*}
 \quad\int_0^1\!\!\!\dif z_1\!\!\int\!\!\du\!&\left\{\Theta(\frac{1}{2}-z_1)
  \P_{\pg\pg}(z_1)R(z_1,\ku,\qq)
  -\frac{\Theta(q-z_1 k_1)}{z_1\,\ku^2\,\qq^2}\right\}=\\
 =\int_0^1\!\!\!\dif z_1\!\!\int\!\!\du\!&\left\{\pp(z_1)R(z_1,\ku,\qq)
 -\frac{1}{\ku^2\qq^2}\frac{1}{z_1}+\frac{1}{\ku^2\qq^2}\log\frac{k_1}{q}\,
 \Theta_{k_1q}\right\}\quad,
\end{align*}
where we have changed variables $z_1\to1-z_1$ and $-\ku\to\ku+\kd$ in
the intermediate steps. The
function $\pp(z_1)$ has only one pole at $z_1=0$ which is correctly
subtracted by the $1/z_1$ term. Decomposing $R(z_1,\ku,\qq)$ in a
sum of terms with only two factors in the denominator, and performing
the $\ku$ integrations as explained for the quark impact factor, we
get, at ${\cal O}(\e)$,
\begin{align}
 \hspace{-8mm}&\frac{\sqrt{\pi}\,2C_A\as N_{\e}}{\sqrt{N_c^2-1}\mu^{2\e}}
 \frac{\ab}{\Gamma(1-\e)\mu^{2\e}}\int_0^1\!\!\dif z_1\!\int\!\!\du
 \left\{\Theta(\frac{1}{2}-z_1)\P_{\pg\pg}(z_1)R(z_1,\ku,\qq)
 -\frac{\Theta(q-z_1 k_1)}{z_1\,\ku^2\,\qq^2}\right\} \nonumber\\
 \hspace{-7mm}&=\!\sqrt{\frac{\pi}{N_c^2-1}}
 \frac{2C_A\as N_{\e}}{\kd^2\mu^{2\e}}
 \frac{\ab\,\Gamma^2(1+\e)}{\e\,\Gamma(1+2\e)}
 \!\!\left(\frac{\kd^2}{\mu^2}\right)^{\e}\!\!\!\!
 \int_0^1\!\!\!\dif z_1\!\!\left[\pp(z_1)\left(1+z_1^{2\e}+(1-z_1)^{2\e}\right)
  -\frac{2}{z_1}+\frac{1}{2\e}\right] \nonumber\\
 \hspace{-7mm}&=h_{\pg}^{(0)}(\kd)\,\o^{(1)}(\kd^2)
 \left[-\frac{2}{\e}+\frac{11}{2}-\left(\frac{67}{9}-
 \frac{2\pi^2}{3}\right)\e\right]\quad.      \label{hgc}
\end{align}

Summing Eqs.~(\ref{hga}), (\ref{hgb}) and (\ref{hgc}) we finally
obtain the one-loop correction to the gluon impact factor:
\begin{align*}
 \hspace{-1mm}&\ab\,h_{\pg}^{(1)}(\kd)=
 \int\du\int_0^1\dif z_1\,F_{\pg}^{(1)}(z_1,\ku,\kd)\\
 &\hspace{-1mm}\;=h_{\pg}^{(0)}(\kd)\o^{(1)}(\kd^2)\!\left[\!\left(
 \frac{11}{6}-\frac{n_f}{3N_c}\right)+
 \left(\frac{11}{6}+\frac{(2+\e)n_f}{6N_c}\right)-\frac{n_f\,C_F}{N_c^2}
 \left(\frac{1}{3}+\frac{1}{6}\e\right)-\frac{\K}{N_c}\e\right]
\end{align*}
as given in Eq.~(\ref{hunog}) of the text.


\end{document}